\documentclass{jpsj2}
\title{%
	Average Structures of a Single Knotted Ring Polymer
}
\author{%
	Shinya {\sc Saka}\thanks{E-mail address: ssaka@rk.phys.keio.ac.jp}
	and
	Hiroshi {\sc Takano}
}
\inst{%
	Faculty of Science and Technology,
	Keio University, Yokohama 223-8522, Japan
}
\abst{%
Two types of average structures of a single knotted ring polymer
are studied by Brownian dynamics simulations.
For a ring polymer with $N$ segments,
its structure is represented by a $3N$-dimensional conformation vector
consisting of
the Cartesian coordinates of the segment positions
relative to the center of mass of the ring polymer.
The average structure is given by
the average conformation vector, which is self-consistently
defined as
the average of the conformation vectors obtained from a simulation
each of which is rotated to minimize
its distance from the average conformation vector.
From each conformation vector sampled in a simulation,
$2N$ conformation vectors
are generated by changing the numbering of the segments.
Among the $2N$ conformation vectors,
the one closest to the average conformation vector
is used
for
one type of the average structure.
The other type of the averages structure
uses all the conformation vectors generated from 
those sampled in a simulation.
In the case of the former average structure,
the knotted part of the average structure
is delocalized for small $N$ and becomes localized
as $N$ is increased.
In the case of the latter average structure,
the average structure changes
from a double loop structure for small $N$
to a single loop structure for large $N$,
which indicates the localization-delocalization transition
of the knotted part.
} 
\kword{
	knot, 
	ring polymer, 
	single polymer, 
	average structure,
	Brownian dynamics simulations, 
	knot localization 
}
\begin{document}
\maketitle
\section{Introduction}%
\label{sec:Introduction}
One of the important problems of the polymer physics is
the effect of the topological constraints on polymer systems.
In the case of systems of linear polymers,
the topological constraints caused by the entanglement of the polymers
vary temporally
and the time-averaged properties of the constraints
have been studied.%
\cite{deGennes,Doi-Edwards}
In contrast,
in the case of a single ring polymer,
the topological constraints caused by its entanglement with itself,
which is called a knot,
are determined by the type of the knot and
do not change with time.
Therefore, a single ring polymer system can be considered
as an ideal system for the study of topological effects
and its investigation is expected to provide
a basis for further understanding of the topological effects
on polymer systems.
The interest in single knotted ring polymers has been growing
in recent years.%
\cite{  Rensburg1991,
	Orlandini1998,
	Shimamura2002,
	Matsuda2003,
	Saka2008,
	Metzler2002,
	Hanke2003,
	Orlandini2003,
	Marcone2005,
	Marcone2007,
	Virnau2005,
	Frago2002
	}
A crucial but unresolved problem in a single knotted ring polymer is
how topological effects of the knot depend on the polymer length.
There have been studies predicting that
the topological effects vanish and properties of the knotted ring
polymer become the same as those of the unknotted ring polymer
for infinitely long chains.%
\cite{  Rensburg1991,
	Orlandini1998,
	Shimamura2002,
	Matsuda2003,
	Saka2008}
This prediction
can be understood from
the argument that
the knotted part of a ring polymer
becomes localized to a part of the polymer
as the polymer length increases and
that the rest of the polymer behave like an unknotted ring polymer.%
\cite{
	Saka2008,
	Metzler2002,
	Hanke2003,
	Orlandini2003,
	Marcone2005,
	Marcone2007,
	Virnau2005,
	Frago2002
	}
The localization of the knotted part has been studied
by measuring the chain length
$l$
of the knotted part
through simulations.%
\cite{Orlandini2003, Marcone2005, Marcone2007, Virnau2005, Frago2002}
It has been found that 
the equilibrium average $\left\langle l \right\rangle$ of $l$ behaves as
$\left\langle l \right\rangle
\propto N^t$  with $t \simeq 0.75$
for a single knotted ring polymer of $N$ segments
in good solvent.%
\cite{Marcone2005, Marcone2007}
The fact that $0 < t < 1$
indicates the localization of knotted part.
It has been discussed
and shown to be consistent with the simulation data
that
the behavior of $\langle l \rangle$ appears
in the $N$-dependence of the mean square radius of gyration
as a leading correction to its scaling behavior.
In contrast, 
a behavior
$\left\langle l \right\rangle \propto N$,
which corresponds to the knot delocalization,
has been found
for a single knotted ring polymer in poor solvent.
Similar results has been obtained for
single ring polymers in two dimensions
\cite{ Metzler2002, Hanke2003, Orlandini2003}
and
linear polymers.
\cite{Virnau2005}
Recently, the present authors studied
the relaxation rate spectrum of single ring homopolymers
in good solvent.
\cite{ Saka2008}
Because
a ring homopolymer has
the translational symmetry along the polymer,
its relaxation rates are classified by a wave number $q$.
In ref.\ \citen{Saka2008},
the relaxation rate $\lambda_q$ for a wave number $q$ is estimated
from 
the time $t$ dependence of the correlation function
$
	C_{q} \left( t \right)
	=
	N^{-1} 
	\sum_{i} \sum_{j}
	\left( 1/3 \right)
	\left\langle
	\left(
		{\mib r}_{i}     \left( t \right) - 
		{\mib r}_{\rm c} \left( t \right)
	\right)
	\cdot
	\left(
		{\mib r}_{j}     \left( 0 \right) -
		{\mib r}_{\rm c} \left( 0 \right)
	\right)
	\right\rangle
	{\rm exp} \left[ -{\rm i} 2\pi q (i-j) / N \right]
$
on the basis of the relaxation mode analysis,%
\cite{Takano1995, Koseki1997, Hirao1997}
where
$
	{\mib r}_{i} \left( t \right)
$
denotes the position of the $i$th segment
of a ring polymer with $N$ segments at time $t$
and
$
	\mib{r}_{\rm c} \left( t \right)
	 = \frac{1}{N} \sum_{i=1}^{N} \mib{r}_i \left( t \right)
$
is the position
of the center of mass of the polymer.
It is found
for single ring polymers with the trefoil knot
that the slowest relaxation rate for each $N$ is given by
$\lambda_q$ with $q=2$ for small values of $N$
and
that with $q=1$ for large values of $N$.
This transition is considered to correspond to
the change of the structure of the ring polymer
caused by the localization of the knotted part.
In the studies mentioned above, however, 
the structures of single knotted ring polymers in three dimensions
are not observed directly.
\par
The purpose of the present paper is
to confirm the localization of the knotted part
of a single knotted ring polymer
by directly observing its average structure in three dimensions
through simulations.
The structure of a ring polymer with $N$ segments
is represented by a $3N$-dimensional conformation vector
consisting of
the Cartesian coordinates of the segment positions
relative to the center of mass of the ring polymer.
The average structure is given by
the average conformation vector, which is self-consistently
defined as
the average of the conformation vectors obtained from a simulation
each of which is rotated to minimize
its distance from the average conformation vector.%
\cite{ McLachelan1979 } 
The average structure has been frequently used
in simulation studies of biopolymers.
Especially,
static properties of
fluctuations from the average structure
have been studied by the principal component analysis method.%
\cite{Kitao1991, Abagyan1992, Garcia1992} 
Recently,
the dynamic properties, such as the relaxation modes and rates,
are also studied by the relaxation mode analysis method.%
\cite{Mitsutake}
In the case of the studies of biopolymers,
polymers studied are usually heteropolymers.
In contrast, 
a ring polymer studied in the present paper is a homopolymer
and has the translational symmetry along the polymer chain.
In the present paper,
by utilizing the translational symmetry,
we propose an extension of
the definition of the average structure
and examine
how the localization of the knotted part appears in
the new and conventional types of the average structures.
\par
The present paper is organized as follows.
In \S2,
a model used in the present study and
the two types of the average structures of 
single homopolymers are explained.
The results of the simulations are presented in \S3.
Summary and discussion are given in the last section.
\section{Model and Average Structures}%
\label{sec:Model}
In order to study a single knotted ring polymer in good solvent,
Brownian dynamics simulations of a bead-spring model are performed.
The dynamics of a single ring polymer
with $N$ segments is described by the Langevin equation
\begin{equation}
	\label{eq:Langevin Eq.}
	\frac{ {\rm d} \mib{r}_i (t) }{ {\rm d} t}
	=
	- \frac{1}{\zeta}
	  \frac{ {\rm d} V \left( \left\{ \mib{r}_j \right\} \right) }
	       { {\rm d} \mib{r}_i }
	+ \mib{w}_i(t).
\end{equation}
Here, 
$\mib{r}_i(t)$ is a three-dimensional column vector
consisting of the Cartesian coordinates of
the position of the $i$th segment
at time $t$ and $\zeta$ is the friction constant.
The random force $\mib{w}_i$ acting on the $i$th segment
is a Gaussian white stochastic process satisfying
$	\langle w_{i,\alpha}(t) \rangle = 0 $
and
$
	\left\langle w_{i,\alpha}(t)w_{j,\beta}(t') \right\rangle
	=
	\left( 2 k_{{\rm B}} T / \zeta \right)
	\delta_{i,j}\delta_{\alpha,\beta}\delta(t-t'),
$
where $w_{i,\alpha}, k_{{\rm B}}$ and $T$ denote
the $\alpha$-component of $\mib{w}_i$, the Boltzmann constant and
the temperature of the system, respectively.
The potential $V(\{ \mib{r}_i \}) = V( \mib{r}_1, \ldots, \mib{r}_N)$
represents the interaction between the segments.
In the present paper,
we use the potential given by%
\cite{Hirao1997,Saka2008,Grest1986,Binder}
\begin{equation}
	\label{eq:potential}
	V(\{ \mib{r}_j \})
	=  \sum_{i=2}^{N} \sum_{j=1}^{i-1}
			V_{{\rm R}}(|\mib{r}_{i} - \mib{r}_{j}|)
	 + \sum_{i=1}^{N} V_{{\rm A}}(|\mib{r}_{i+1} - \mib{r}_{i}|),
\end{equation}
where
$\mib{r}_{N+1} = \mib{r}_1$ in the last summation of the right-hand side
because
the $N$th segment is connected to the first segment.
Here, $V_{{\rm R}}$ is given by
the repulsive part of the Lennard-Jones potential
\begin{equation}
	\label{eq:potential LJ}
	V_{{\rm R}}(r)
	=
	\left\{ 
	\begin{array}{lll}
		\displaystyle
			4 \epsilon \Bigl[ 
				 \Bigl( \frac{\sigma}{r} \Bigr)^{12}
				-\Bigl( \frac{\sigma}{r} \Bigr)^{6}
				+\frac{1}{4}
				\Bigr]
			&{\rm for}& r \leq 2^{\frac{1}{6}}\sigma, \\
		0
			&{\rm for}& r >    2^{\frac{1}{6}}\sigma,
	\end{array}
	\right.
\end{equation}
and represents the excluded volume
interaction between all the segments.
The potential $V_{{\rm A}}$, which is called
a finitely extensible nonlinear elastic (FENE) potential, is given by
\begin{equation}
	\label{eq:potential FENE}
	V_{{\rm A}}(r)
	=
	\left\{ 
	\begin{array}{lll}
		\displaystyle
		-\frac{1}{2}k R_0^2 
		 \ln \Bigl[ 1-\Bigl( \frac{r}{R_0} \Bigr)^2 \Bigr]
				&{\rm for}& r < R_0, \\
		\infty
				&{\rm for}& r \ge  R_0,
	\end{array}
	\right.
\end{equation} 
and represents the attractive interaction
between neighboring segments along the ring polymer.
The same model is used
for simulations of a single linear polymer with $N$ segments,
where the upper limit of the last summation of
the right-hand side of eq.\ (\ref{eq:potential})
is $N-1$.
\par
In the following, the average structures
are explained for a linear or ring homopolymer with $N$ segments.
The average structures are estimated by using
conformations of the polymer sampled in simulations.
In a simulation,
$M$ conformations are sampled at interval of time ${\mit \Delta T}$
after an initial equilibration time $T_{\rm i}$.
Let $\mib{R}_i(m)$
denote the position of the $i$th segment
relative to the center of mass of the polymer
in the $m$th sample:
\begin{equation}
	\mib{R}_i(m) = \mib{r}_i (t_m) - \mib{r}_{\rm c}(t_m),
\end{equation}
and
$
t_m = T_{\rm i} +(m-1) {\mit \Delta} T
$, where $m = 1, \ldots, M$.
A structure of the polymer in the $m$th sample is represented by
a $3N$-dimensional conformation vector
$
\mib{C}(m)
=
{}^{\rm t}\!\!%
\left(
{}^{\rm t}\!%
\mib{R}_1(m), 
{}^{\rm t}\!%
\mib{R}_2(m), 
\ldots,  
{}^{\rm t}\!%
\mib{R}_N(m)
\right)
$.
From each $\mib{C}(m)$,
$K$ conformations, $\tilde{\mib{C}}(m,k)$ with $k=1, \ldots, K$,
each of which has the same statistical weight as $\mib{C}(m)$,
can be generated
by changing the numbering of the segments,
which is possible for homopolymers.
Here, $K = 2$ for linear homopolymers and $K=2N$ for ring homopolymers.
In both cases,
$\tilde{\mib{C}}(m,k)$ is given by
$
\tilde{\mib{C}}(m,k) =
{}^{\rm t}\!\!%
\left(
{}^{\rm t}\!%
\mib{R}_{n(1,k)}(m),
{}^{\rm t}\!%
\mib{R}_{n(2,k)}(m),
\ldots,
{}^{\rm t}\!%
\mib{R}_{n(N,k)}(m)
\right)
$,
where
$n(i,k)= (1-2r)i_j+r(N+1)$
with
$i_j = ((i+j-1) \mod N)+1$,
$j = \lfloor (k-1)/2 \rfloor$,
and
$r = (k-1) \mod 2 = k-1 -  2 j$.
Note that $i_j$ represents the integer between $1$ and $N$
which is equal to $i+j$ modulo $N$.
For odd or even values of $k$,
$n(i,k)$ is equal to $i_j$ or $N - i_j + 1$, respectively.
\par
From the set of conformation vectors
$\{\tilde{\mib{C}}(m,k); m=1,\ldots,M ~\mbox{and}~ k=1,\ldots,K\}$,
the average conformation vector 
$
\mib{C}^{{\rm av}\ast}
=
{}^{\rm t}\!\!%
\left(
{}^{\rm t}\!%
\mib{R}_1^{{\rm av}\ast},
{}^{\rm t}\!%
\mib{R}_2^{{\rm av}\ast},
\ldots,
{}^{\rm t}\!%
\mib{R}_N^{{\rm av}\ast}
\right)
$, which gives
one of the two types of
the average structures, is calculated as follows.
Here, the superscript av$\ast$ denotes either av0 or av1
represeinting the type-0 or type-1 average structure
explained in the following,
respectively.
For each conformation vector $\tilde{\mib{C}}(m,k)$,
a three-dimensional rotation $\mathcal{R}_{m,k}$ 
which minimizes the square distance
$d^2(m,k)$
between the conformation vector
$\mathcal{R}_{m,k}\left(\tilde{\mib{C}}(m,k)\right)$
obtained by the rotation of $\tilde{\mib{C}}(m,k)$
and the average conformation vector 
$\mib{C}^{{\rm av}\ast}$
is determined.\cite{ McLachelan1979 } 
Here, 
$
\mathcal{R}_{m,k}\left(\tilde{\mib{C}}(m,k)\right)
=
{}^{\rm t}\!\!%
\left(
{}^{\rm t}%
\mathcal{R}_{m,k}\left(\mib{R}_{n(1,k)}(m)\right),
{}^{\rm t}%
\mathcal{R}_{m,k}\left(\mib{R}_{n(2,k)}(m)\right),
\ldots,
{}^{\rm t}%
\mathcal{R}_{m,k}\left(\mib{R}_{n(N,k)}(m)\right)
\right)
$
and
\begin{equation}
d^2(m,k)
=
\left(\mathcal{R}_{m,k}\left(\tilde{\mib{C}}(m,k)\right)
-\mib{C}^{{\rm av}\ast}\right)^2 =
\sum_{i=1}^{N}
\left(\mathcal{R}_{m,k}\left(\mib{R}_{n(i,k)}(m)\right)
- \mib{R}_i^{{\rm av}\ast}\right)^2{}.
\end{equation}
Thus,
the rotation $\mathcal{R}_{m,k}$ fits 
the conformation $\tilde{\mib{C}}(m,k)$ to
the average conformation $\mib{C}^{{\rm av}\ast}$.
\par
In the case of the conventional average structure,
which has been used for heteropolymers,
the $i$th segment of a sampled structure is fitted to 
the $i$th segment of the average structure.
We call this type of average structure
the type-0 average structure (type-0 AS).
Because the numbering of the segments is crucial,
the conformation vectors
$\tilde{\mib{C}}(m,k)$, $k=1,\ldots,K$
are considered to represent different structures,
although they are obtained from $\mib{C}(m)$
by only changing the numbering of the segments.
The average conformation vector $\mib{C}^{\rm av0}$
for the type-0 AS
is then given by
\begin{equation}
\mib{C}^{\rm av0}
=
\frac{1}{MK}\sum_{m=1}^M \sum_{k=1}^K
\mathcal{R}_{m,k}\left( \tilde{\mib{C}}(m,k) \right).
\label{av0}
\end{equation}
\par
In the following,
we propose a new type of an average structure
for homopolymers,
which we call the type-1 average structure (type-1 AS).
Because all the segments of a homopolymer are equivalent,
the conformation vectors $\tilde{\mib{C}}(m,k)$, $k=1,\ldots,K$
are considered to represent the same structure as $\mib{C}(m)$,
although they have different numbering of the segments.
Therefore the best fit of a sampled conformation $\mib{C}(m)$ to
the average conformation
is given by
the rotation $\mathcal{R}_{m,k_{\rm min}(m)}$,
where $k_{\rm min}(m)$ is the value of $k$ which gives 
the smallest value of $d^2(m,k)$ with $m$ fixed.
Thus, 
the average conformation vector $\mib{C}^{\rm av1}$
for the type-1 AS
is given by
\begin{equation}
\mib{C}^{\rm av1}
=
\frac{1}{M}\sum_{m=1}^M
\mathcal{R}_{m,k_{\rm min}(m)}
\left( \tilde{\mib{C}}(m,k_{\rm min}(m)) \right).
\label{av1}
\end{equation}
\par
Because the definition of the rotation $\mathcal{R}_{m,k}$
contains $\mib{C}^{{\rm av}\ast}$,
eqs.\ (\ref{av0}) and (\ref{av1}) should be solved self-consistently.
In practice, $\mib{C}^{{\rm av}\ast}$ is calculated iteratively.
By using
the $n$th candidate for the average conformation vector
$\mib{C}^{{\rm av}\ast}_n$
calculated from the $n$th iteration, which may be chosen as
one of $\tilde{\mib{C}}(m,k)$
for the first iteration ($n=0$),
all the rotations
$\mathcal{R}_{m,k}$ are determined.
Then,
the right-hand side of eq.\ (\ref{av0}) or (\ref{av1})
is calculated and the result is used as
the next candidate $\mib{C}^{{\rm av}\ast}_{n+1}$.
The calculation is iterated for $n=0, 1, 2, \ldots$
until
the distance between
$\mib{C}^{{\rm av}\ast}_n$ and $\mib{C}^{{\rm av}\ast}_{n+1}$ becomes
sufficiently small
and the self-consistency of eq.\ (\ref{av0}) or (\ref{av1})
is achieved.
\par
\section{Results}
\label{sec:Result and Discussion}
\par
Brownian dynamics simulations
of the model described in the previous section
are performed
for a single linear polymer and 
single ring polymers with the trivial knot and the trefoil knot.
The following parameters are used:%
\cite{Grest1986, Hirao1997, Saka2008}
$k_{{\rm B}}T/\epsilon=1, k\sigma^2/\epsilon = 30$
and $R_{{\rm 0}}/\sigma = 1.5$.
The Euler algorithm with a time step
${\mit \Delta} t = 10^{-4} \zeta \sigma^2/\epsilon$
is employed for a numerical
integration of the equation of motion (\ref{eq:Langevin Eq.}).
Hereafter, we set $\sigma=1$, $\zeta=1$ and $\epsilon=1$.
\par
In order to calculate the average structures,
conformations of a single polymer with $N$ segments
are taken from a simulation
every ${\mit \Delta} T = 10^{-2} \tau(N)$.
For single linear polymers,
the time $\tau(N)$
is chosen as
$
	\tau( N ) = A_{{\rm L}} N^{x_{\rm L}}
$,
which corresponds to the behavior of 
the longest relaxation time of single linear polymers.%
\cite{deGennes,Doi-Edwards,Koseki1997,Hirao1997}
For single ring polymers,
$\tau(N)$ is chosen as
$
	\tau ( N ) = A_{{\rm R}} N^{x_{\rm R}}
$,
which corresponds to the behavior of 
the longest relaxation time for the wave number $q=1$
of single ring polymers with the trivial knot.%
\cite{Saka2008}
The parameters are chosen as
$A_{\rm L} = 21.05$, $x_{\rm L}=2.22$,
$A_{\rm R} = 22.96$ and $x_{\rm R}=2.095$.
The values for linear polymers are
estimated from the relaxation mode analysis of
the time-displaced correlation matrix
$
	\langle
	 \left( {\mib r}_i(t) - {\mib r}_{\rm c}(t) \right)
	 \cdot
	 \left( {\mib r}_j(0) - {\mib r}_{\rm c}(0) \right)
	 \rangle
$
calculated through simulations.
The values for ring polymers are taken from the previous study.%
\cite{Saka2008}
The equilibration time $T_{\rm i}$ is chosen as
$T_{\rm i} \simeq 10 \tau(N)$.
The number of sampled conformations $M$ is given by
$M = 10^5 $ for $N=30$, $40$, $60$ and $80$,
$M = 2 \times 10^5 $ for $N=110$ and $120$,
$M = 1.5 \times 10^5 $ for $N=160$ and
$M = 5 \times 10^4 $ for $N=240$, respectively.
\par
\begin{figure}[!ht]
\begin{center}
	\includegraphics[height=5.5cm]{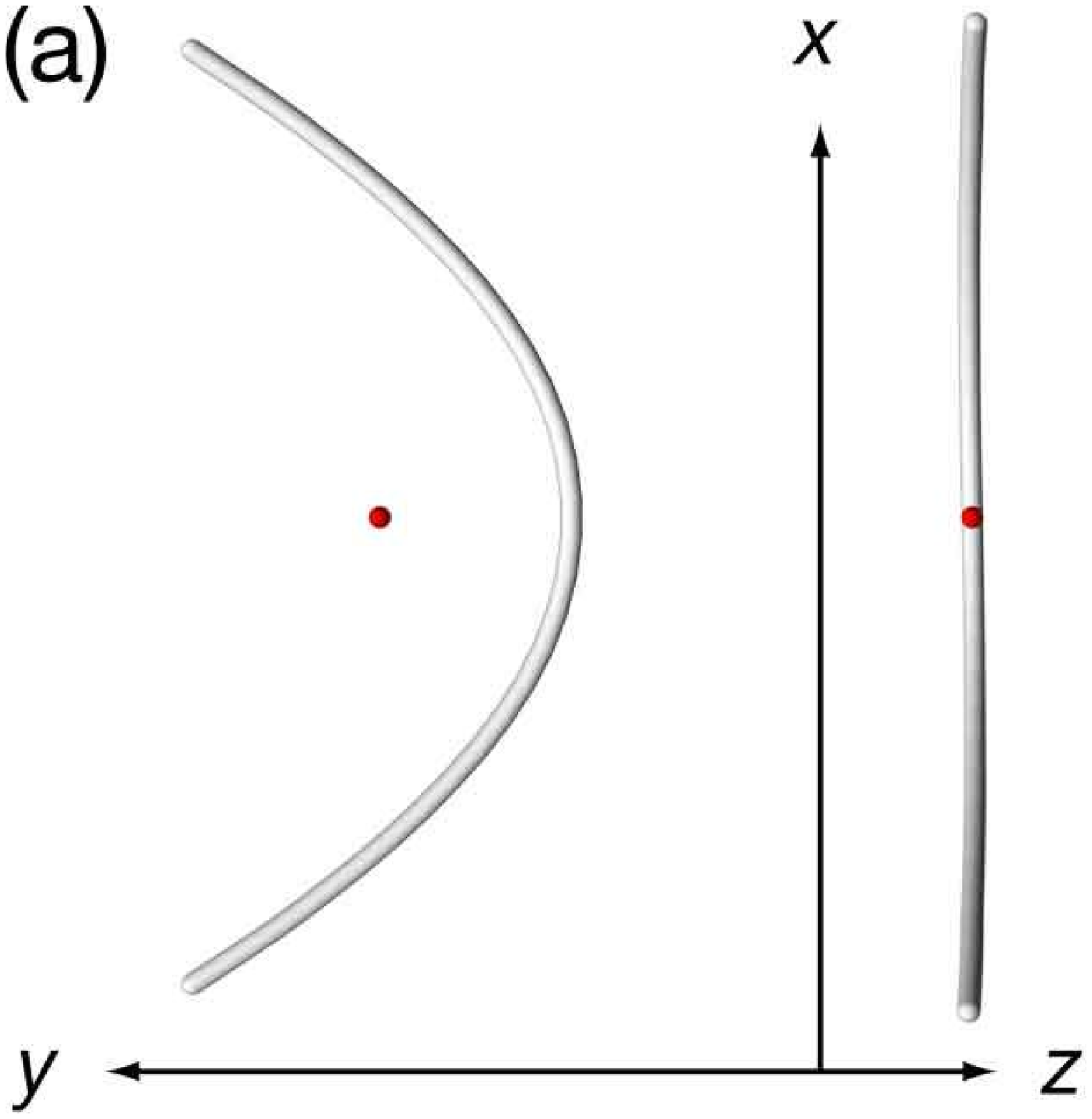}
	\includegraphics[height=5.5cm]{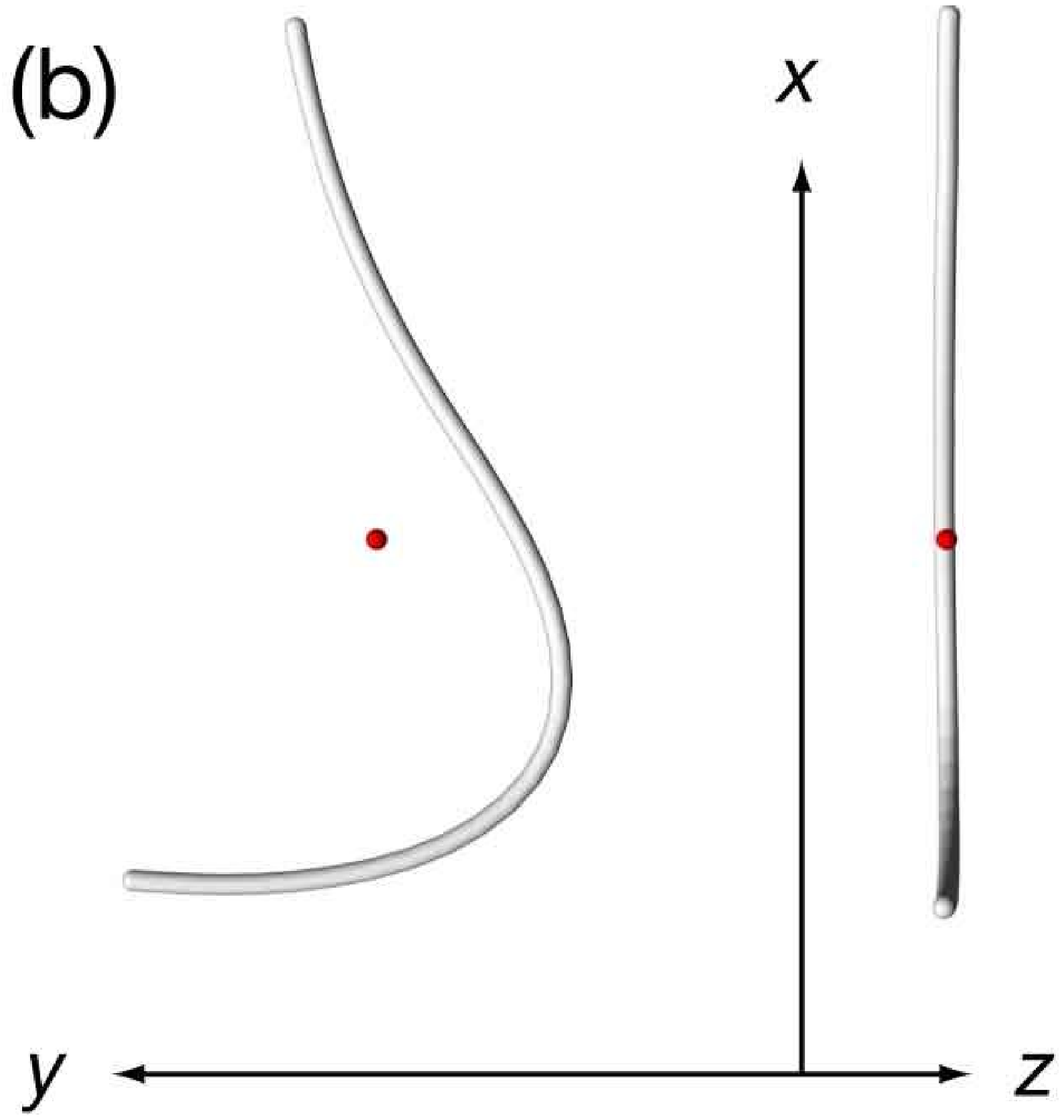}
	\caption{	
		The type-0 AS (a) and the type-1 AS (b)
		of a single linear polymer with $N=40$.
		In each figure,
		the cylinders and the sphere represent
		the bonds connecting adjacent segments
		and the center of mass of the polymer, respectively.
	}
	\label{fig:Average_Structure_Linear}
\end{center}
\end{figure}

Figure \ref{fig:Average_Structure_Linear} shows the average structures
for a single linear polymer with $N=40$ segments.
In this and the following figures,
the bonds connecting adjacent segments of a polymer
are represented by cylinders
and the center of mass of the polymer is shown as a sphere.
The axes of the Cartesian coordinates are chosen
to be the principal axes of the moment of inertia tensor
and the length scale is chosen
to normalize the contour length of each average structure.
In the figures, only the directions of the axes are indicated.
The origin of the Cartesian coordinates is chosen to
be the center of mass of each average structure
in the following descriptions.
The type-0 AS shown in Fig.\ \ref{fig:Average_Structure_Linear}(a) 
has a parabolic shape in the $x$-$y$ plane,
which has C$_2$ symmetry about the $y$ axis.
This result agrees with that of the previous study
of the average structure of single linear polymers\cite{private}.
The C$_2$ symmetry is induced by the symmetry
of a single linear homopolymer that
the $i$th and ($N-i+1$)th segments are equivalent,
because the numbering of the segments is conserved in
the type-0 AS.
On the other hand,
the type-1 AS shown in Fig.\ \ref{fig:Average_Structure_Linear}(b)
does not have the symmetry,
although it has a similar shape in the $x$-$y$ plane.
Each sampled conformation $\mib{C}(m)$
does not have the C$_2$ symmetry,
that is,
there is no rotation which changes
$\mib{C}(m)=\tilde{\mib{C}}(m,1)$ into $\tilde{\mib{C}}(m,2)$.
This asymmetry in the sampled conformations
is conserved in the type-1 AS,
because it neglects the numbering of the segments.
The reason why the type-1 AS is in the $x$-$y$ plane is as follows.
Let $\mathcal{M}$ denote a reflection operation
with respect to a plane containing the origin,
which is the center of mass of conformations in the present case.
For a rotation operation $\mathcal{R}$,
we define another rotation operation as $\mathcal{R}'=\mathcal{MRM}$.
Then, 
$
\left(
\mathcal{R}\left(\tilde{\mib{C}}(m,k)\right)
-
\mib{C}^{{\rm av}\ast}
\right)^2
=
\left(
\mathcal{M}\left(\mathcal{R}\left(\tilde{\mib{C}}(m,k)\right)\right)
-
\mathcal{M}\left(\mib{C}^{{\rm av}\ast}\right)
\right)^2
=
\left(
\mathcal{R}'\left(\mathcal{M}\left(\tilde{\mib{C}}(m,k)\right)\right)
-
\mathcal{M}\left(\mib{C}^{{\rm av}\ast}\right)
\right)^2
$
holds.
Therefore,
if $\mathcal{R}_{m,k_{\rm min}(m)}$
gives the smallest square distance
calculated from $\mib{C}(m)$ and $\mib{C}^{\rm av1}$,
$\mathcal{R}_{m,k_{\rm min}(m)}'$
gives the smallest square distance
calculated from $\mathcal{M}\left(\mib{C}(m)\right)$
and $\mathcal{M}\left(\mib{C}^{\rm av1}\right)$.
By considering the mirror image of eq.\ (\ref{av1}), we have
\begin{eqnarray}
\mathcal{M}\left(%
\mib{C}^{\rm av1}%
\right)
&=&
\frac{1}{M}\sum_{m=1}^M
\mathcal{M}\left(%
\mathcal{R}_{m,k_{\rm min}(m)}
\left( \tilde{\mib{C}}(m,k_{\rm min}(m)) \right)%
\right)
\notag
\\
&=&
\frac{1}{M}\sum_{m=1}^M
\mathcal{R}_{m,k_{\rm min}(m)}'\left(%
\mathcal{M}\left(%
\tilde{\mib{C}}(m,k_{\rm min}(m))%
\right)%
\right).
\end{eqnarray}
This equation means that
if the type-1 AS calculated from
$M$ samples $\mib{C}(m)$, $m = 1, \ldots, M$ is given by
$\mib{C}^{\rm av1}$
then
that calculated from
$M$ samples $\mathcal{M}\left(\mib{C}(m)\right)$, $m = 1, \ldots, M$
is given by
$\mathcal{M}\left(\mib{C}^{\rm av1}\right)$.
For a sampled conformation $\mib{C}(m)$,
its mirror image $\mathcal{M}\left( \mib{C}(m) \right)$
has the same statistical weight as $\mib{C}(m)$
in the ensemble of conformations,
because a single linear polymer has no chirality.
In other words,
the ensemble of conformations $\mathcal{M}\left(\mib{C}(m)\right)$
is the same as that of conformations $\mib{C}(m)$.
Therefore, the average conformation $\mib{C}^{\rm av1}$
and its mirror image
$\mathcal{M}\left(\mib{C}^{\rm av1}\right)$
represent the same conformation,
that is, 
there exists a rotation $\mathcal{R}$
which satisfies
$
\mib{C}^{\rm av1} =
\mathcal{R}\left(\mathcal{M}\left(\mib{C}^{\rm av1}\right)\right)
$.
It should be a conformation in a plane
or
a three-dimensional conformation with a plane of reflection symmetry.
\par
\begin{figure}[!ht]
\begin{center}
	\includegraphics[height=5.5cm]{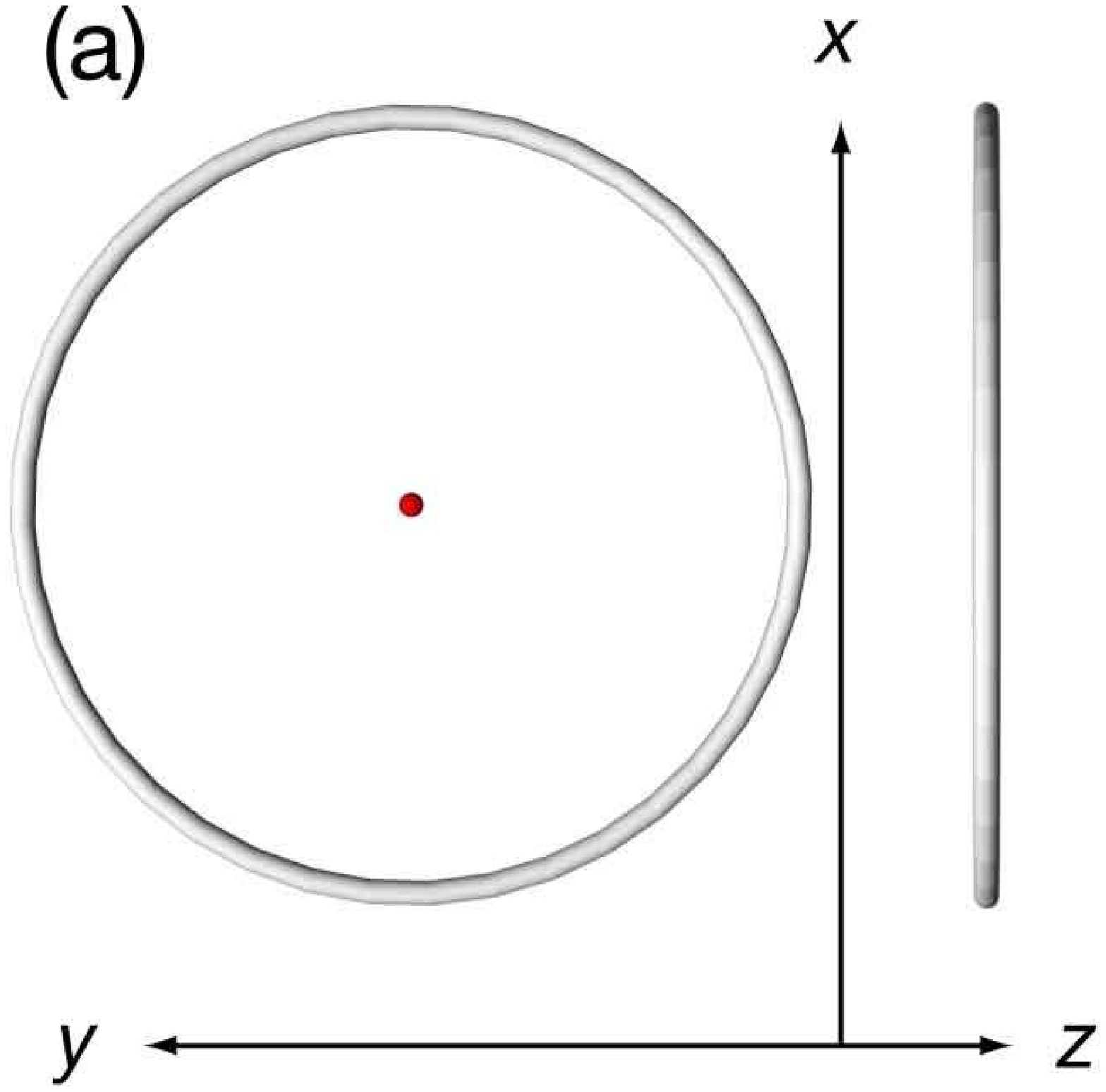}
	\includegraphics[height=5.5cm]{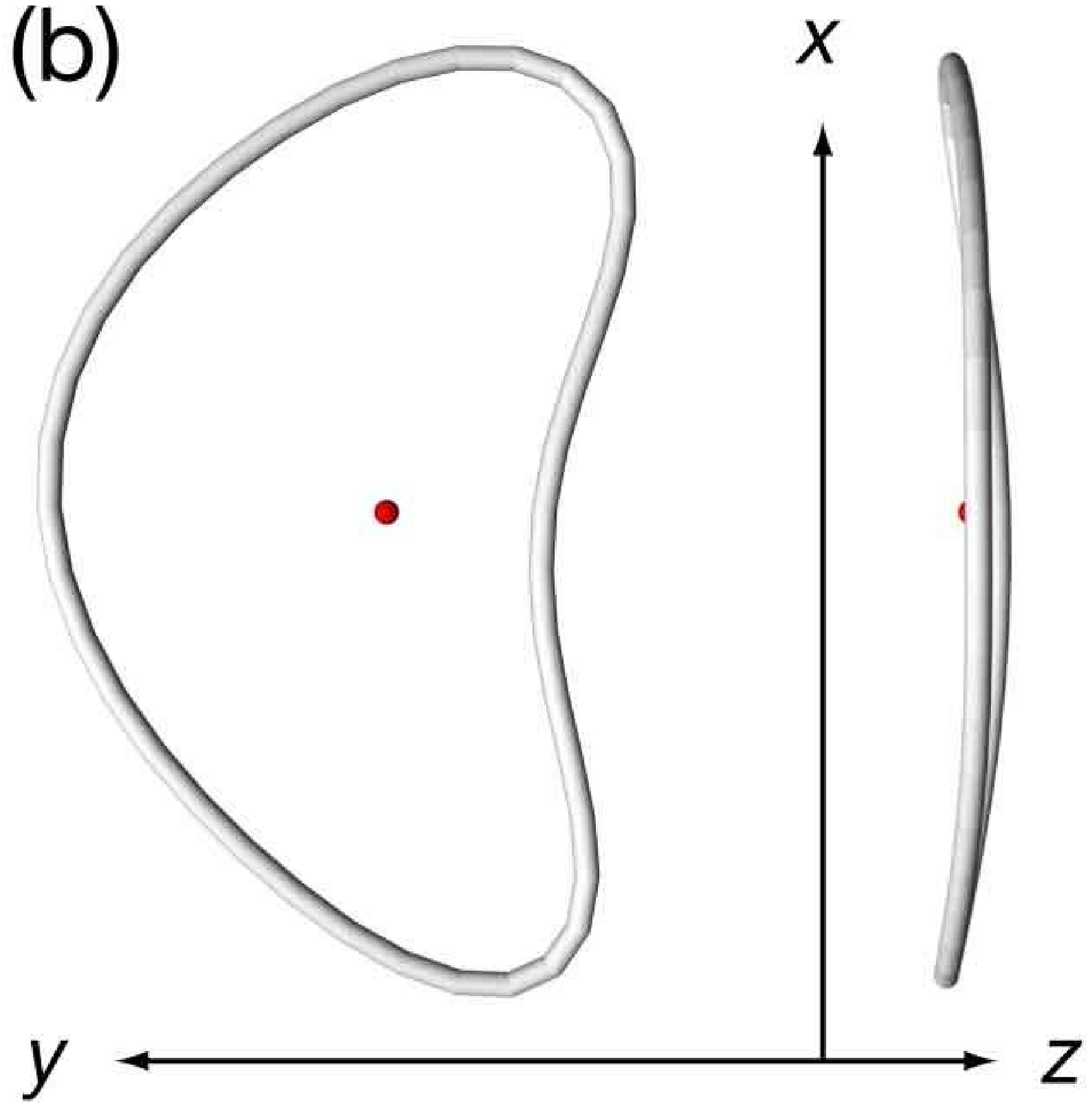}
	\caption{	
		The type-0 AS (a) and the type-1 AS (b)
		of a single ring polymer with the trivial knot
		for $N=40$.
		The meaning of the cylinders and the spheres is
		the same as in Fig.\ \ref{fig:Average_Structure_Linear}.
	\label{fig:Average_Structure_K=O}
	}
\end{center}
\end{figure}

Figure \ref{fig:Average_Structure_K=O} shows
the average structures
of a single ring polymer with the trivial knot
for the case of $N=40$.
The type-0 AS shown in Fig.\ \ref{fig:Average_Structure_K=O}(a)
forms a regular polygon of $N=40$ sides in the $x$-$y$ plane.
As mentioned before,
$2N$ conformations $\tilde{\mib{C}}(m,k)$, $k=1, \ldots, 2N$
are generated from a sampled conformation $\mib{C}(m)$
by changing the numbering of the segments.
The $2N$ ways of changing the numbering correspond
to the $2N$ symmetry operations of the dihedral group D$_N$.
Because all the $2N$ conformations are used with the same 
statistical weight in the calculation of the type-0 AS,
the type-0 AS should be invariant under
the $2N$ ways of changing the numbering.
Therefore, the type-0 AS should be a regular polygon of $N$ sides
in two dimensions,
which has the D$_N$ symmetry,
if all the position vectors in the average structure
are different each other.
In contrast,
the type-1 AS shown in 
Fig.\ \ref{fig:Average_Structure_K=O}(b)
forms a distorted polygon,
which does not have the D$_N$ symmetry,
in the $x$-$y$ plane.
The form of the type-1 AS is explained in the same way as before.
Because the numbering of the segments is neglected in the type-1 AS,
the deviations from the D$_N$ symmetric form
in the sampled conformations
are conserved in the type-1 AS.
Because a sampled conformation and
its mirror image have the same statistical
weight in the ensemble of conformations of a ring polymer with
the trivial knot, which has no chirality,
the average structures should have reflection symmetry.
A structure in a plane is one of the possible structures.
\par
\begin{figure}[!ht]
\begin{center}
	\includegraphics[height=5.5cm]{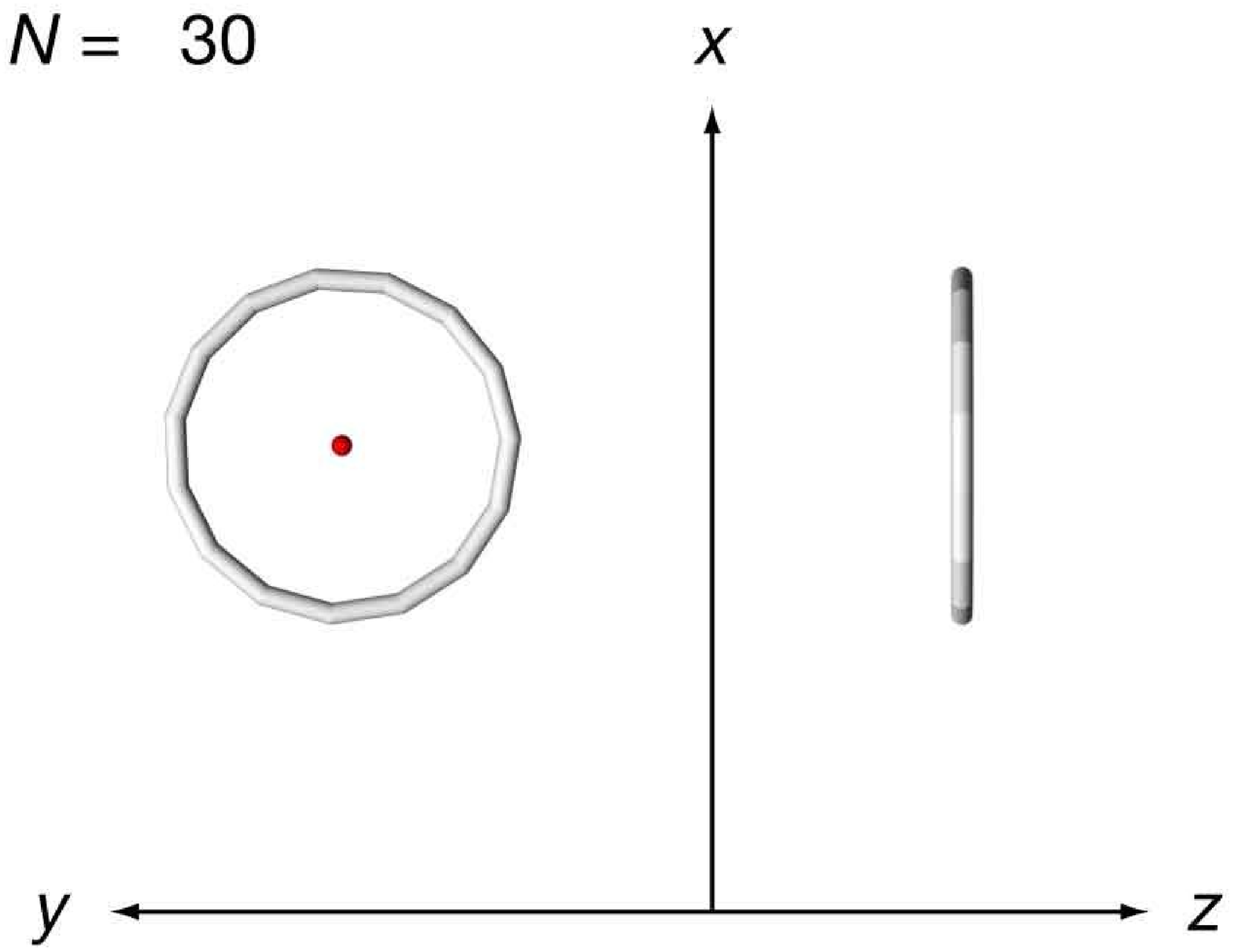}
	\includegraphics[height=5.5cm]{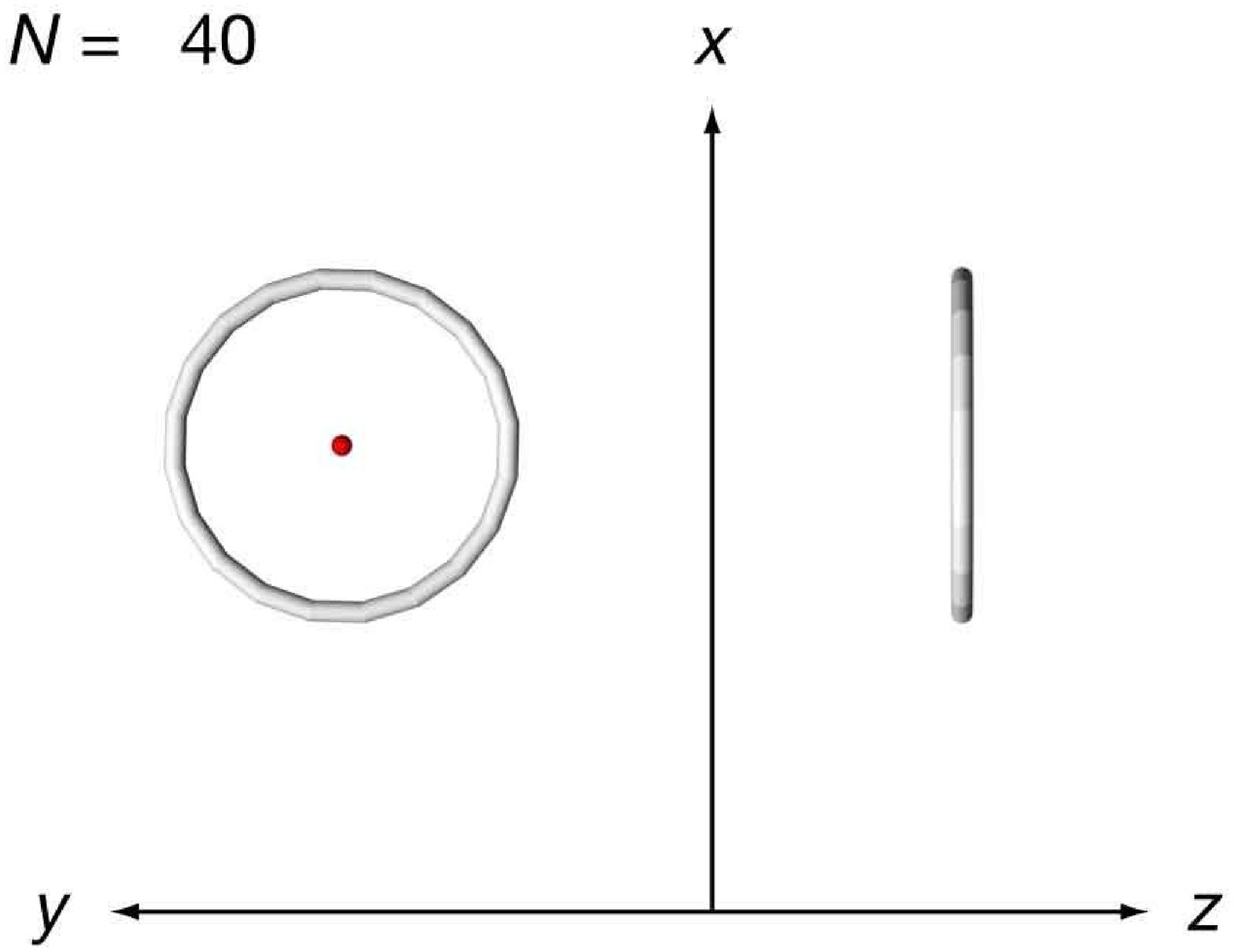}
	\\\medskip
	\includegraphics[height=5.5cm]{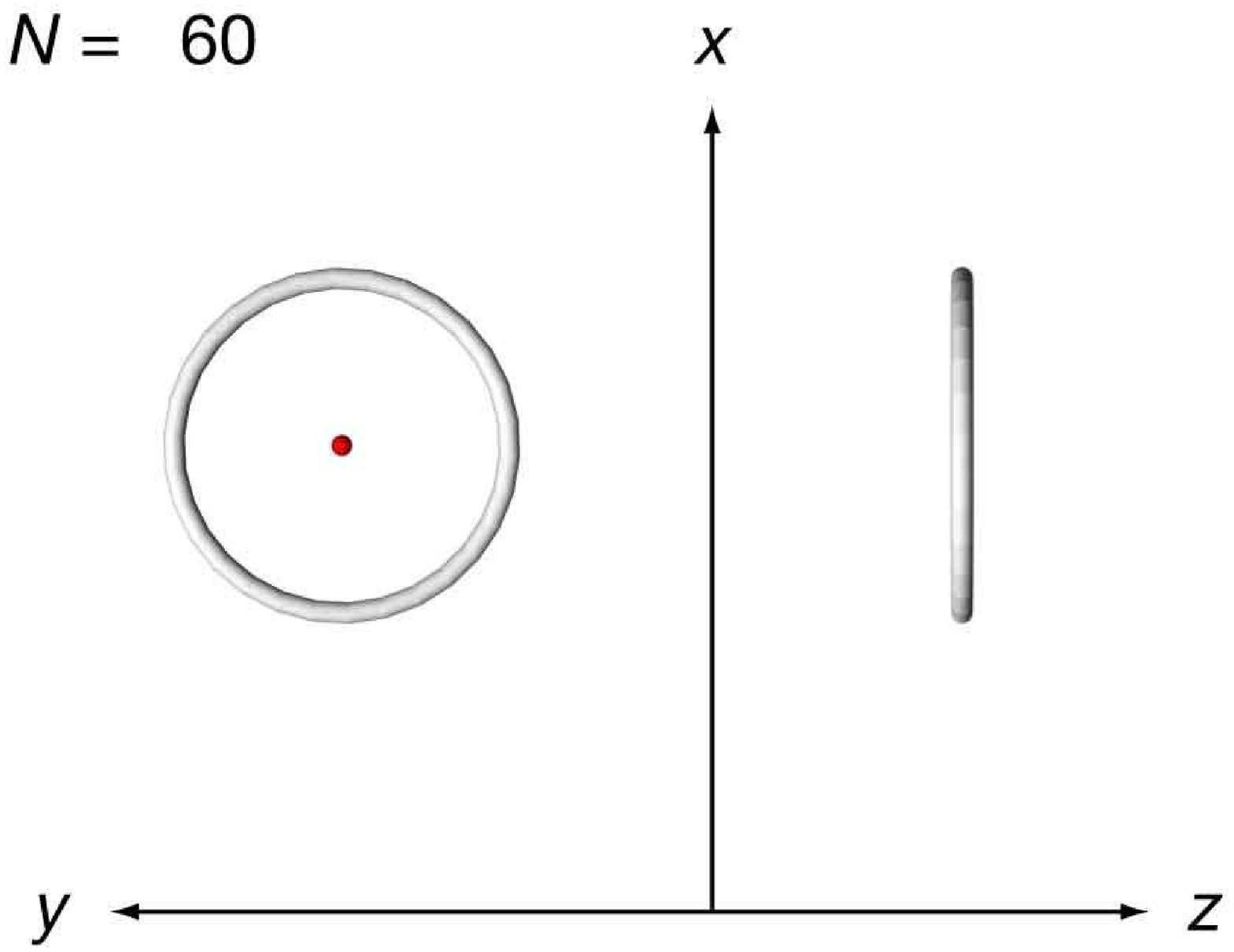}
	\includegraphics[height=5.5cm]{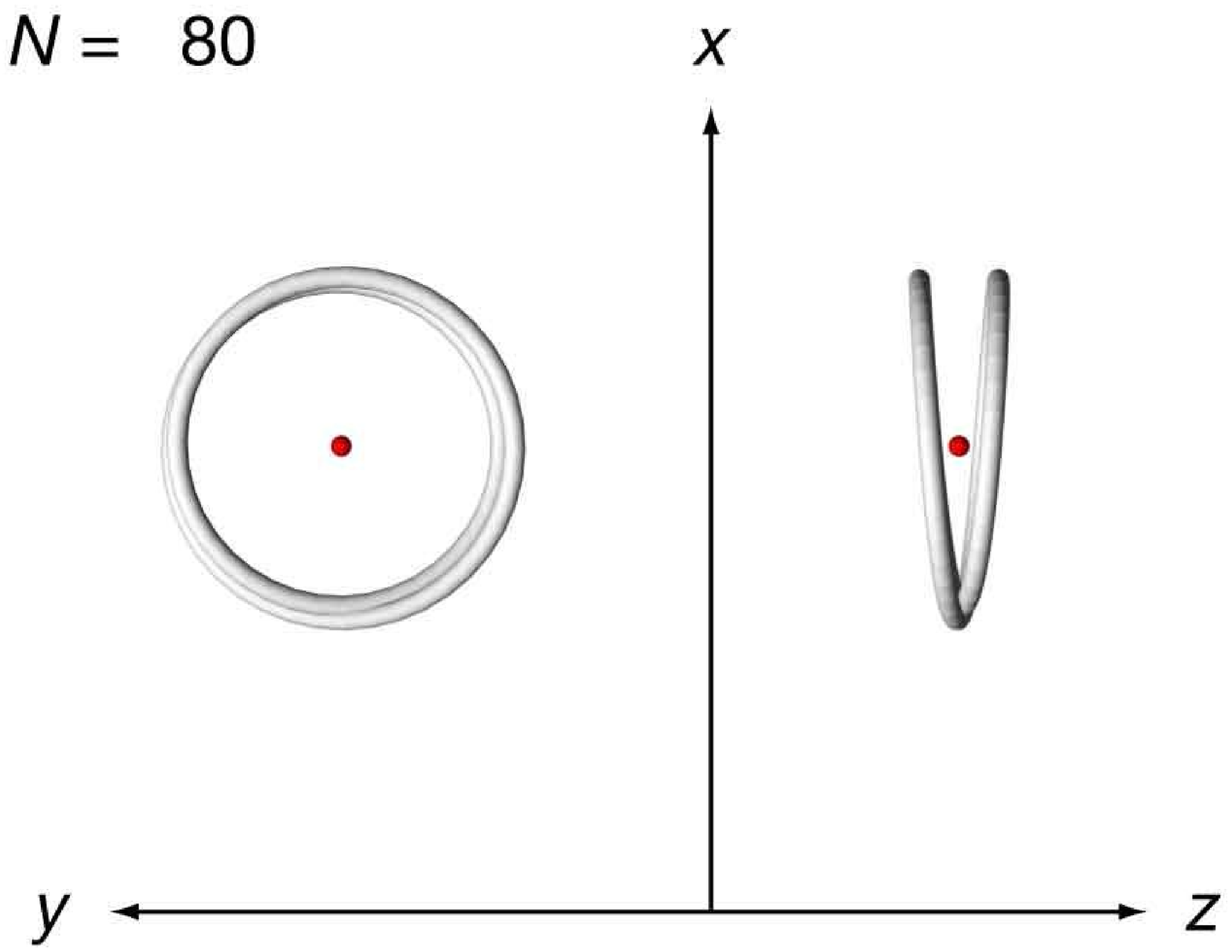}
	\\\medskip
	\includegraphics[height=5.5cm]{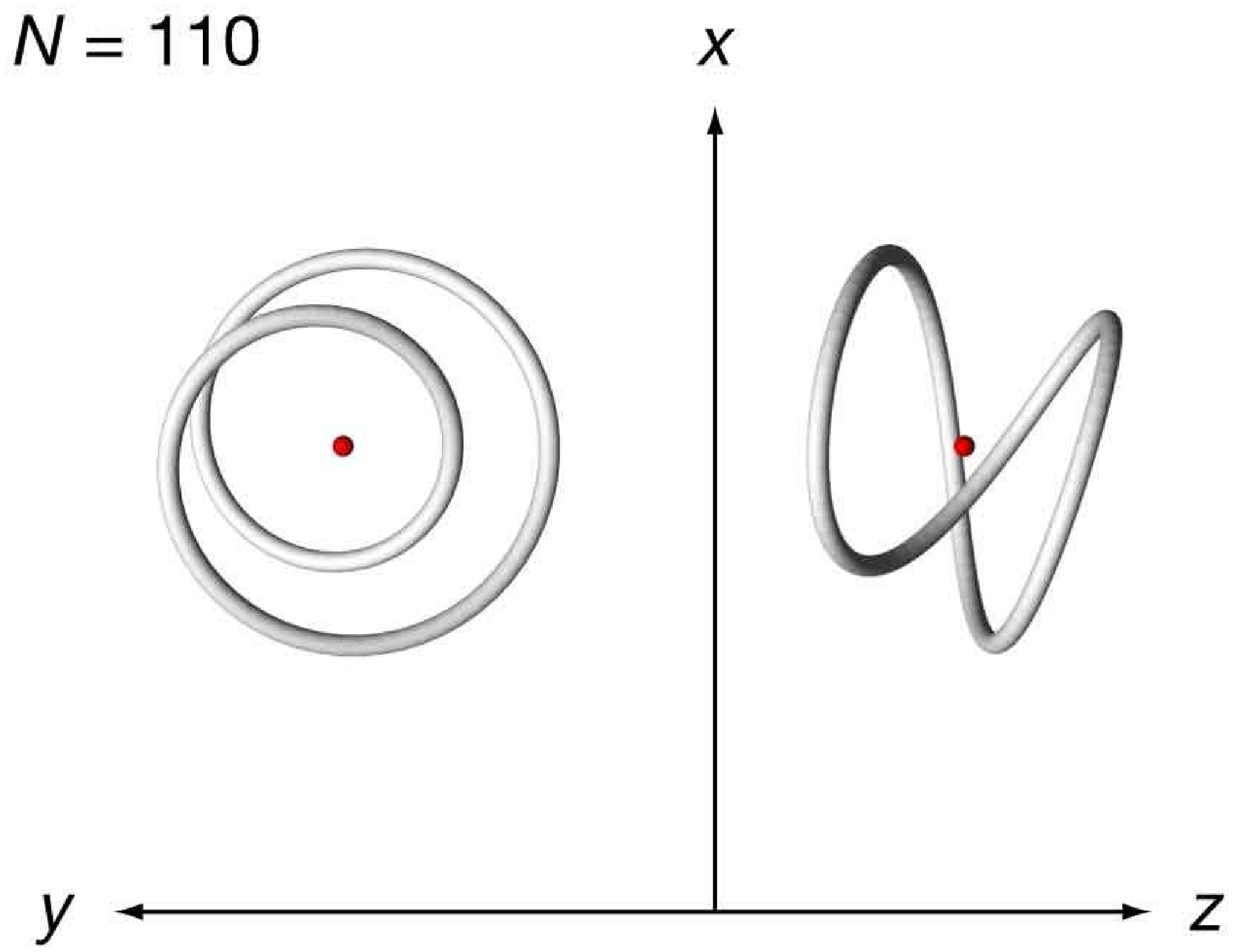}
	\includegraphics[height=5.5cm]{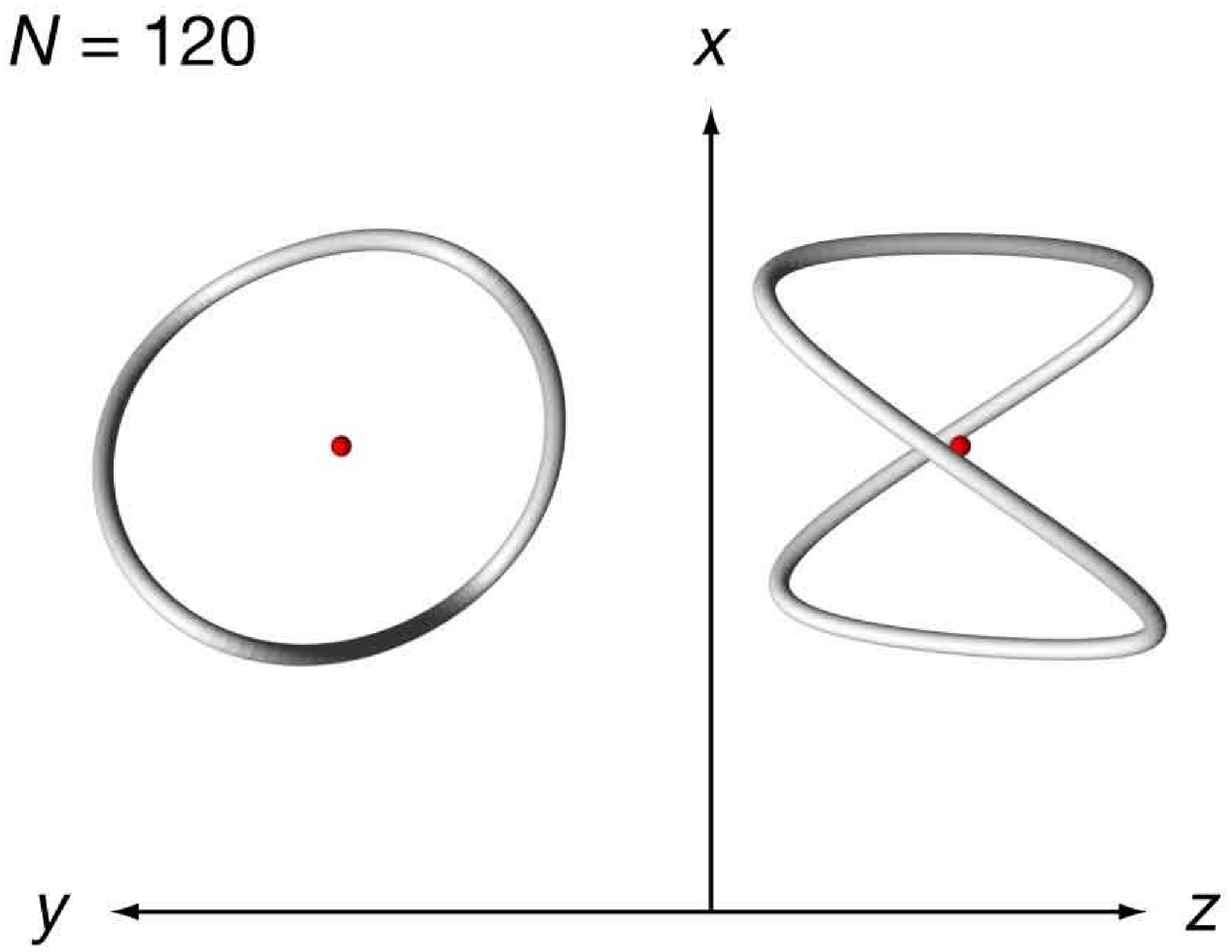}
	\\\medskip
	\includegraphics[height=5.5cm]{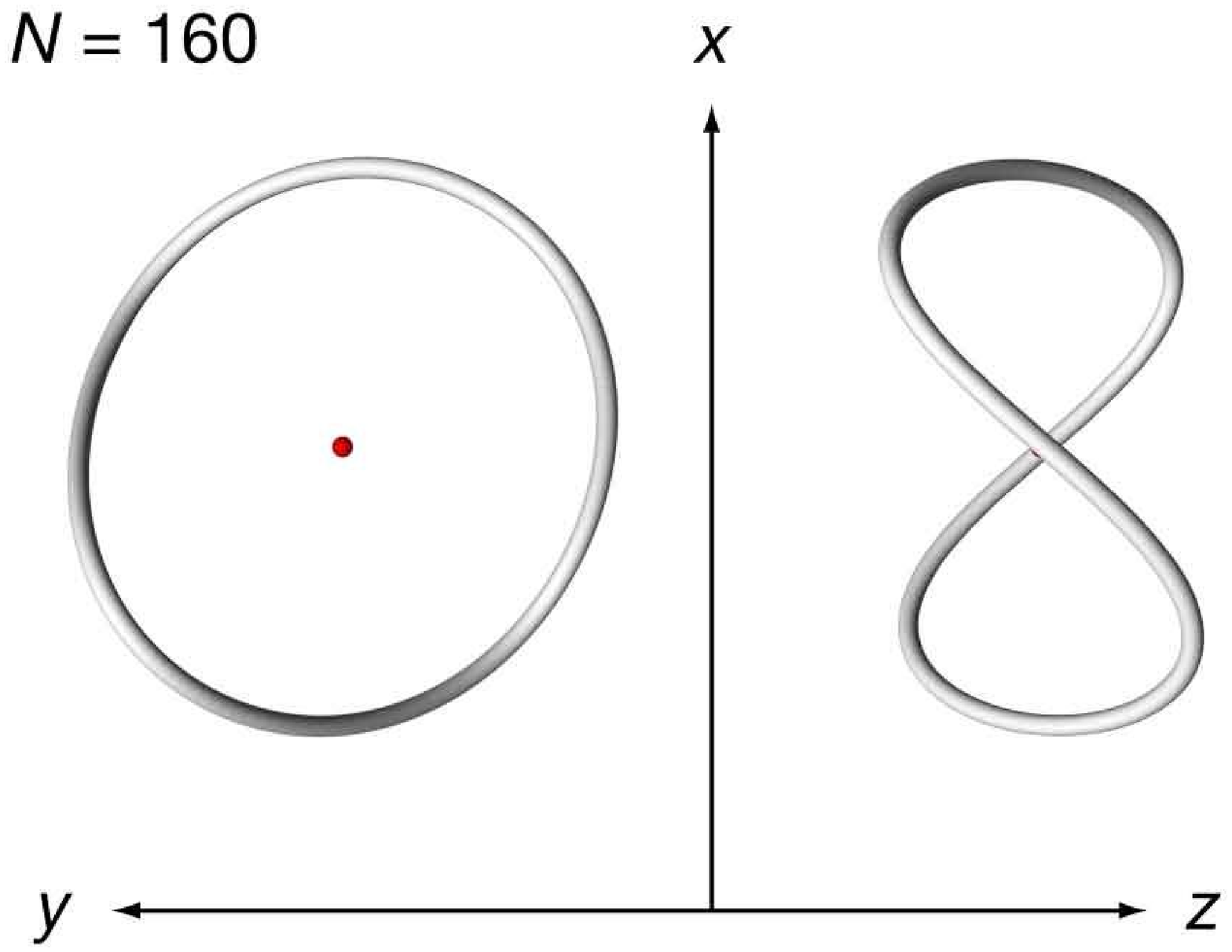}
	\includegraphics[height=5.5cm]{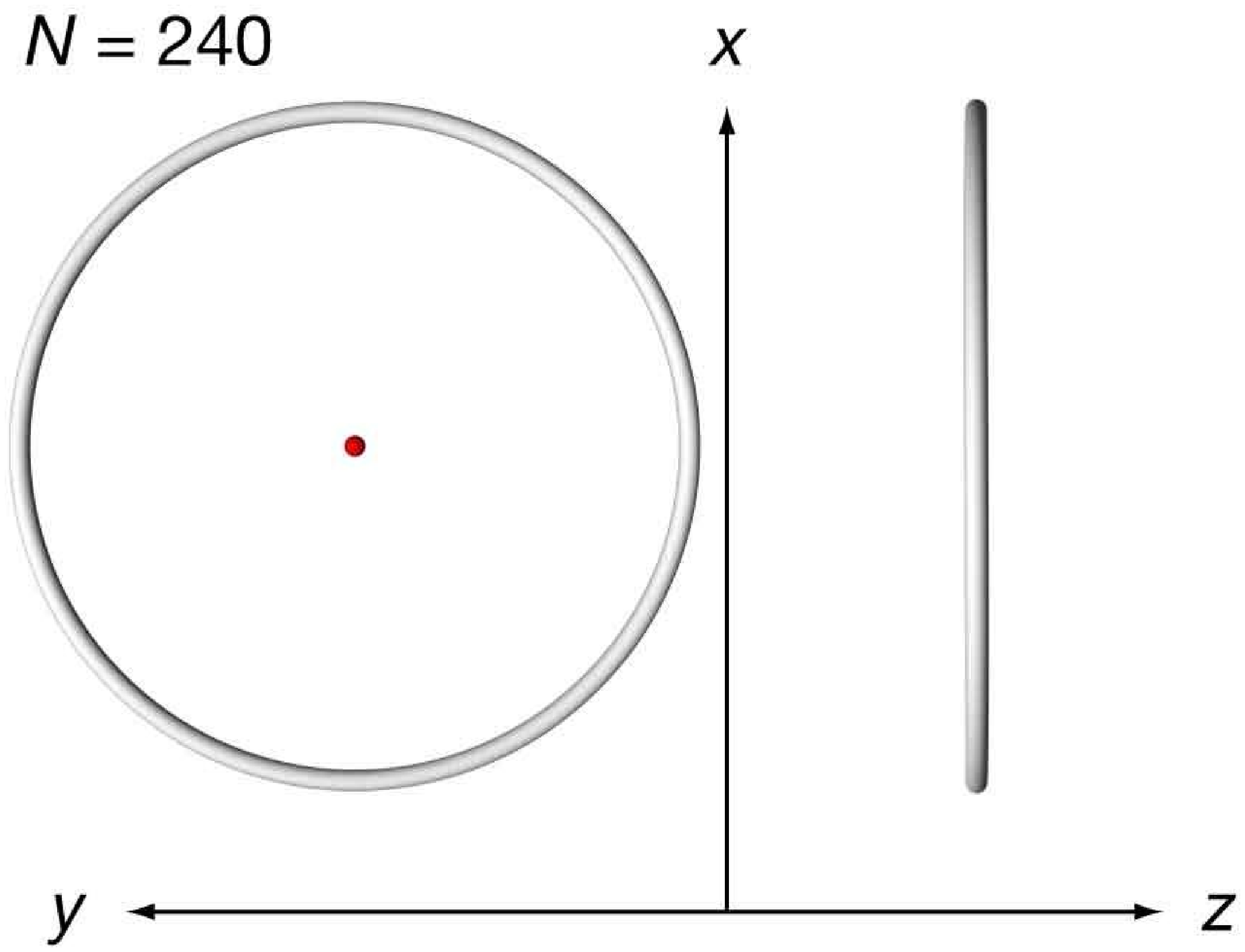}
	\caption{	
		The type-0 ASs
		of a single ring polymer with the trefoil knot
		for $N=30$, $40$, $60$, $80$, $110$, $120$, $160$ and $240$.
		The meaning of the cylinders and the spheres is
		the same as in Fig.\ \ref{fig:Average_Structure_Linear}.
		}
	\label{fig:Average_Structure_K=3_1 AS0}
\end{center}
\end{figure}
\begin{figure}[!ht]
\begin{center}
	\includegraphics[width=7.5cm]{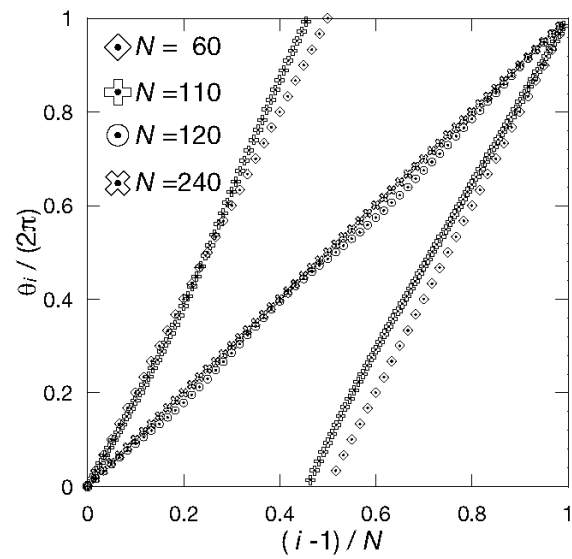}
	\caption{
		A plot of $\theta_i/(2\pi)$ versus $(i-1)/N$
		for the type-0 ASs
		of single ring polymers with the trefoil knot
		for $N=60$, $110$, $120$ and $240$.
		For clarity, only the data for
		$(i-1)/N = n/60$ with $n= 0, 1, \ldots, 59$
		are plotted for $N=120$ and $240$.
	}
	\label{fig:theta}
\end{center}
\end{figure}

Figure \ref{fig:Average_Structure_K=3_1 AS0} shows
the type-0 ASs of single ring polymers
with the trefoil knot for
$N=30$, $40$, $60$, $80$, $110$, $120$, $160$ and $240$.
As explained before,
the type-0 AS should be invariant
under the symmetry operations corresponding to 
the $2N$ ways of changing the numbering of the segments.
Therefore, the deviations from the $x$-$y$ plane
in the average structures for $80 \le N \le 160$ are considered
to be due to statistical errors.
If the deviations for $80 \le N \le 160$ are ignored,
the type-0 ASs shown in  Fig.\ \ref{fig:Average_Structure_K=3_1 AS0}
form regular polygons in the $x$-$y$ plane.
In order to see how the segments are distributed on the regular polygon,
we calculate the angle $\theta_i$ between
the position vectors projected onto the $x$-$y$ plane
of the first and the $i$th segments:
\begin{eqnarray}
	\cos \theta_i &=&
	\frac{
		  R_{1,x}^{{\rm av0}} R_{i,x}^{{\rm av0}}
		+ R_{1,y}^{{\rm av0}} R_{i,y}^{{\rm av0}}
	}
	{
		\sqrt{ {R_{1,x}^{{\rm av0}}}^2 + {R_{1,y}^{{\rm av0}}}^2 }
		\sqrt{ {R_{i,x}^{{\rm av0}}}^2 + {R_{i,y}^{{\rm av0}}}^2 }
	},
	\\
	\sin \theta_i &=&
	\frac{
		  R_{1,x}^{{\rm av0}} R_{i,y}^{{\rm av0}} 
		- R_{1,y}^{{\rm av0}} R_{i,x}^{{\rm av0}}
	}
	{
		\sqrt{ {R_{1,x}^{{\rm av0}}}^2 + {R_{1,y}^{{\rm av0}}}^2 }
		\sqrt{ {R_{i,x}^{{\rm av0}}}^2 + {R_{i,y}^{{\rm av0}}}^2 }
	}.
	\label{eq:theta}
\end{eqnarray}
Here,
$R_{i,\alpha}^{{\rm av0}}$ denotes
the $\alpha$-component of
the position vector ${\mib R}_i^{\rm av0}$ of the $i$th segment
relative to the center of mass in the type-0 AS.
Figure \ref{fig:theta} shows
a plot of $\theta_i/2\pi$ versus $(i-1)/N$.
It is clearly seen that
the type-0 AS forms a double loop for $N \le 110$
and a single loop for $N \ge 120$.
In other words,
the type-0 AS forms
a regular polygon of $N/2$ sides for $N \le 110$,
where $N$ is even,
and 
a regular polygon of $N$ sides for $N \ge 120$.
Note that
in the case of the double loop structure with even $N$,
the symmetry operations corresponding to 
the $2N$ ways of changing the numbering of the segments
become those of D$_{N/2}$,
because $\mib{R}_i^{\rm av0} = \mib{R}_{n(i,N/2)}^{\rm av0}$
for $i = 1, \cdots, N/2$.
The region of the double loop structure
and that of the single loop structure
must be separated by a transition segment number
$N_{\rm t}^{{\rm av0}}$ between $110$ and $120$.
This transition
is considered to correspond to
the localization-delocalization transition of the knotted part,
because
if the knotted part of a conformation 
$\mathcal{R}_{m,k}\left( \tilde{\mib{C}}(m,k) \right)$
is delocalized,
its projection onto the plane of the type-0 AS
encompasses the center of mass
while the projection of the knotted part
does not encompass the center of mass
if the knotted part is localized.
Note that
the form of
the type-0 AS of the ring polymer with the trefoil knot
for $N \ge 120$, which is a regular polygon of $N$ sides,
is the same
as that with the trivial knot.
\par
\begin{figure}[!ht]
\begin{center}
	\includegraphics[height=5.5cm]{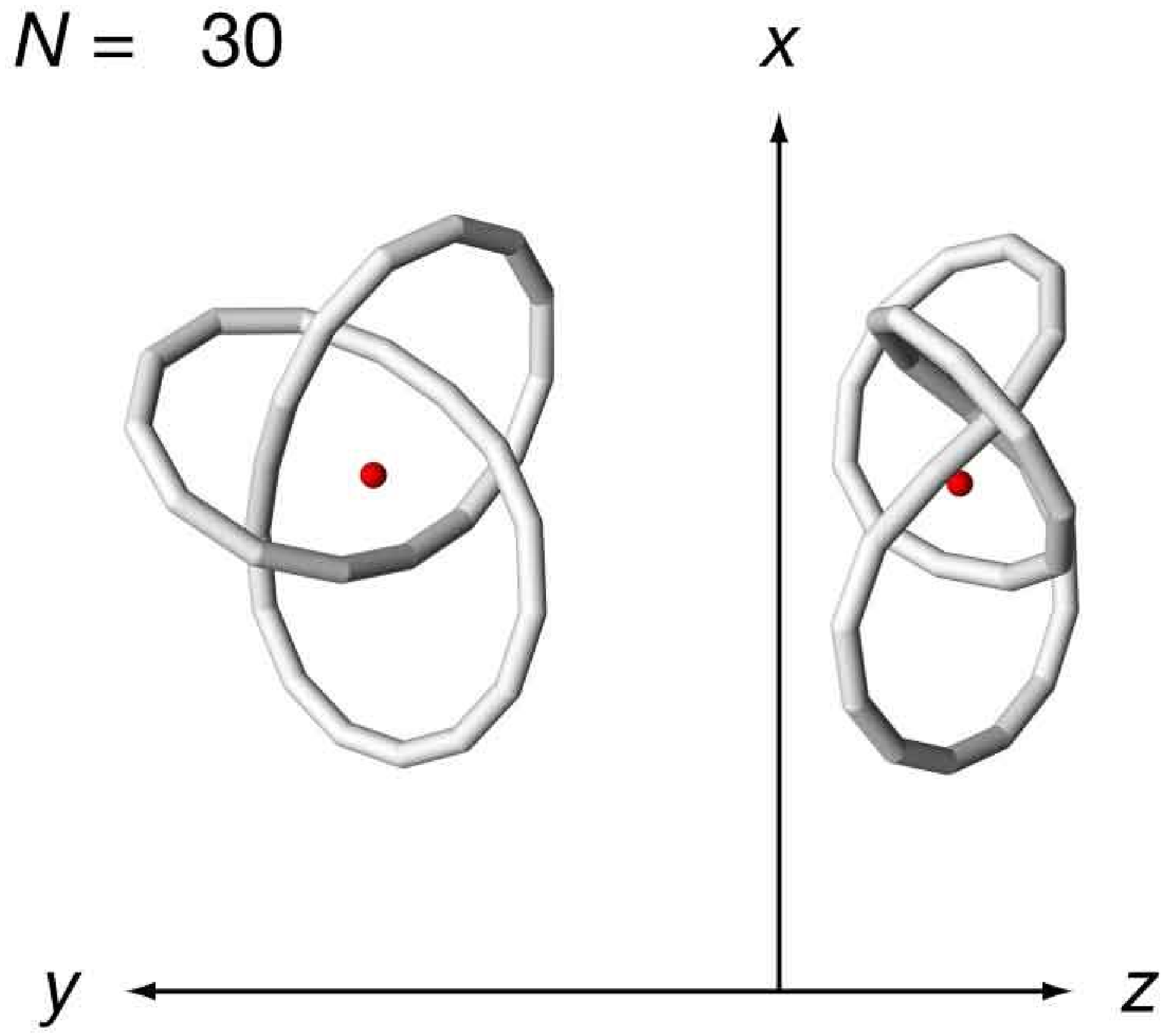}
	\includegraphics[height=5.5cm]{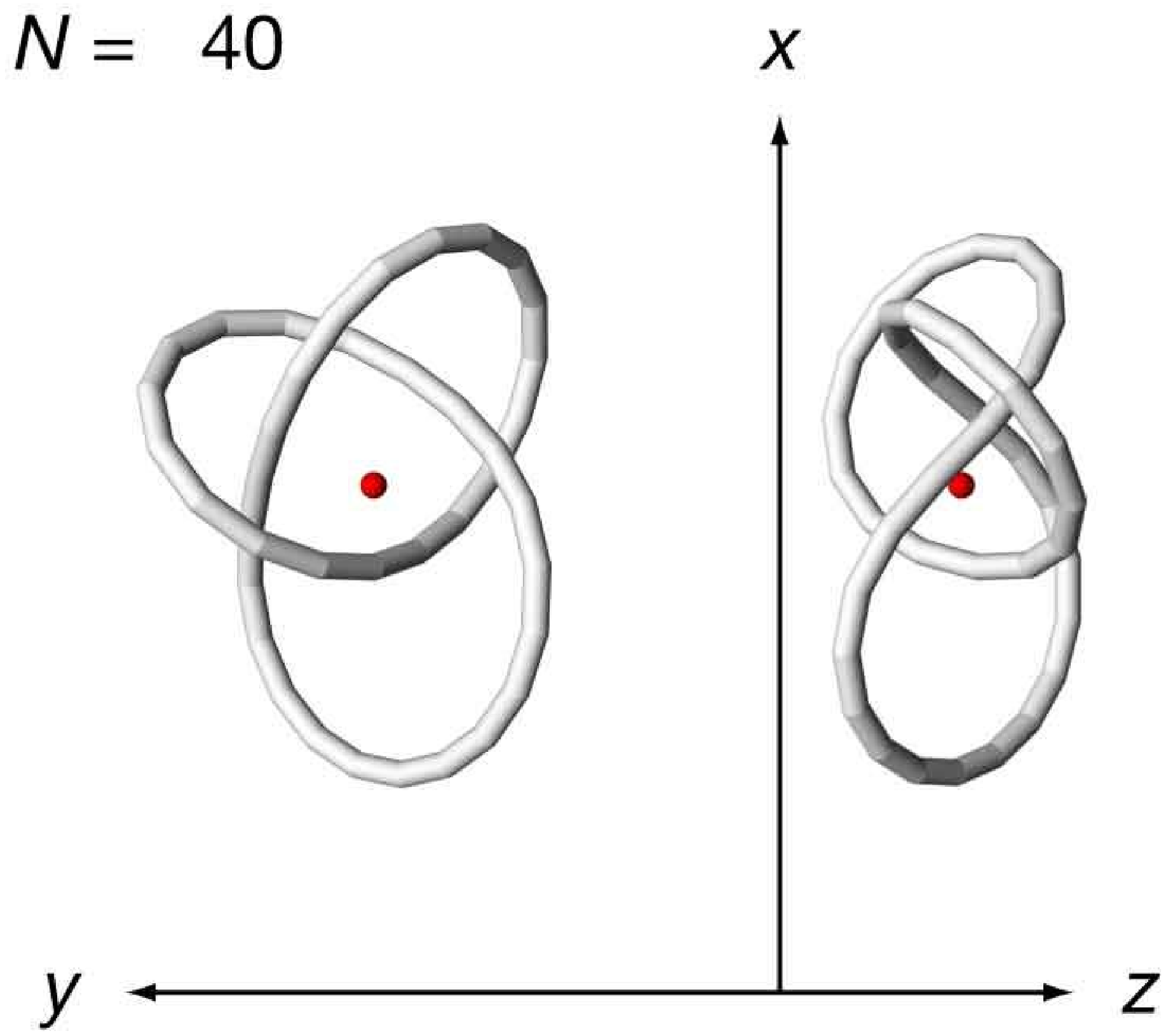}
	\\\medskip
	\includegraphics[height=5.5cm]{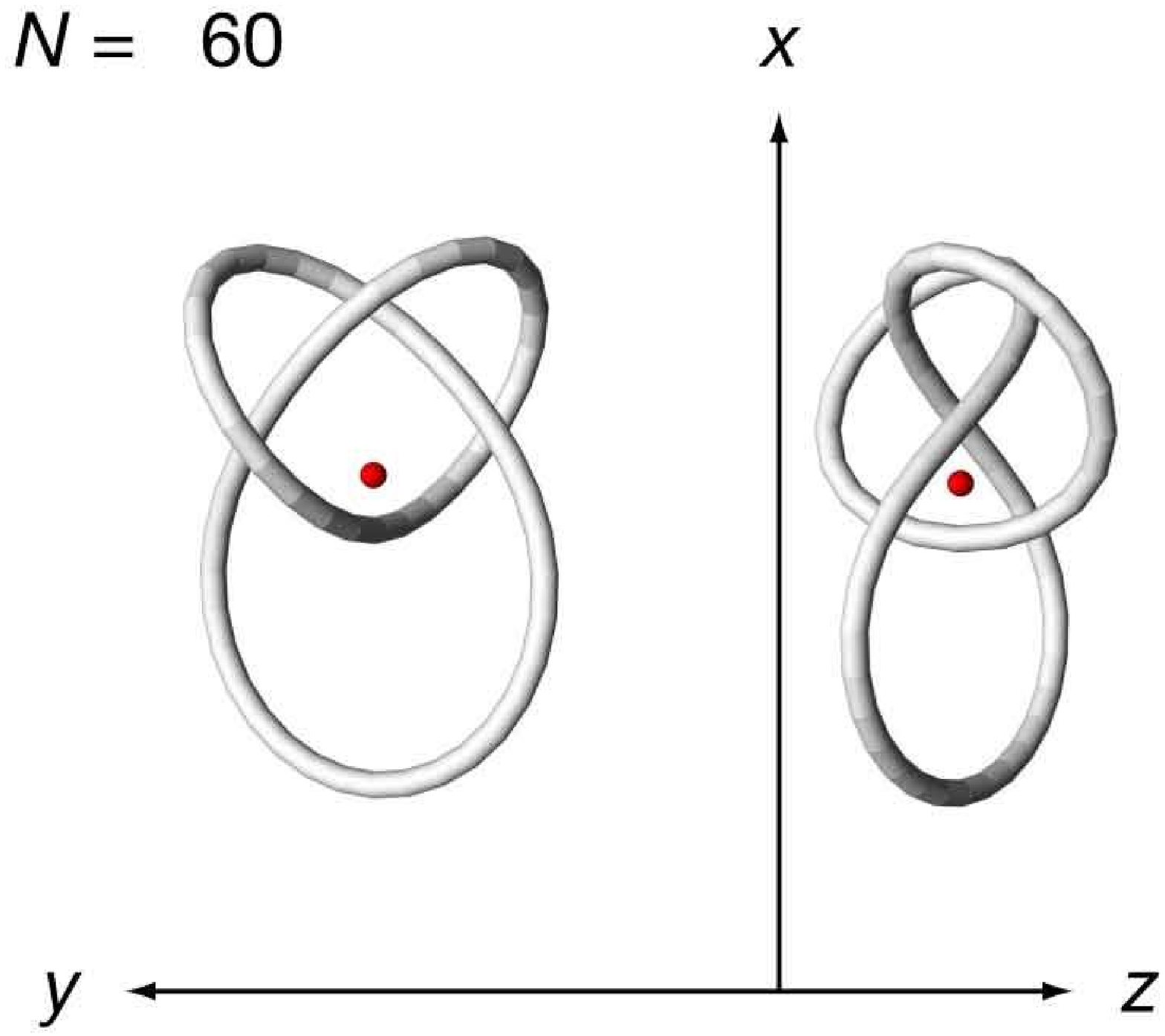}
	\includegraphics[height=5.5cm]{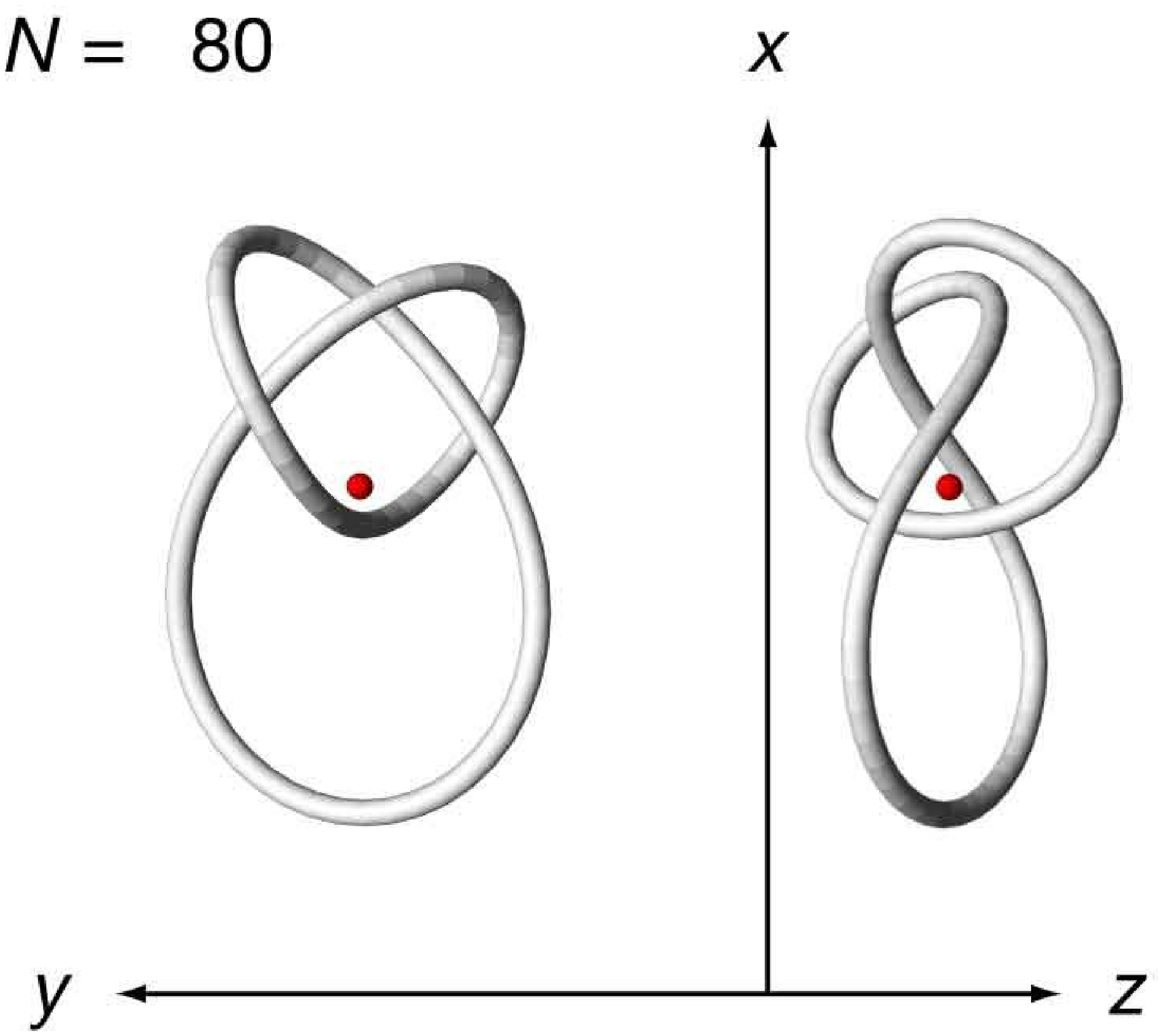}
	\\\medskip
	\includegraphics[height=5.5cm]{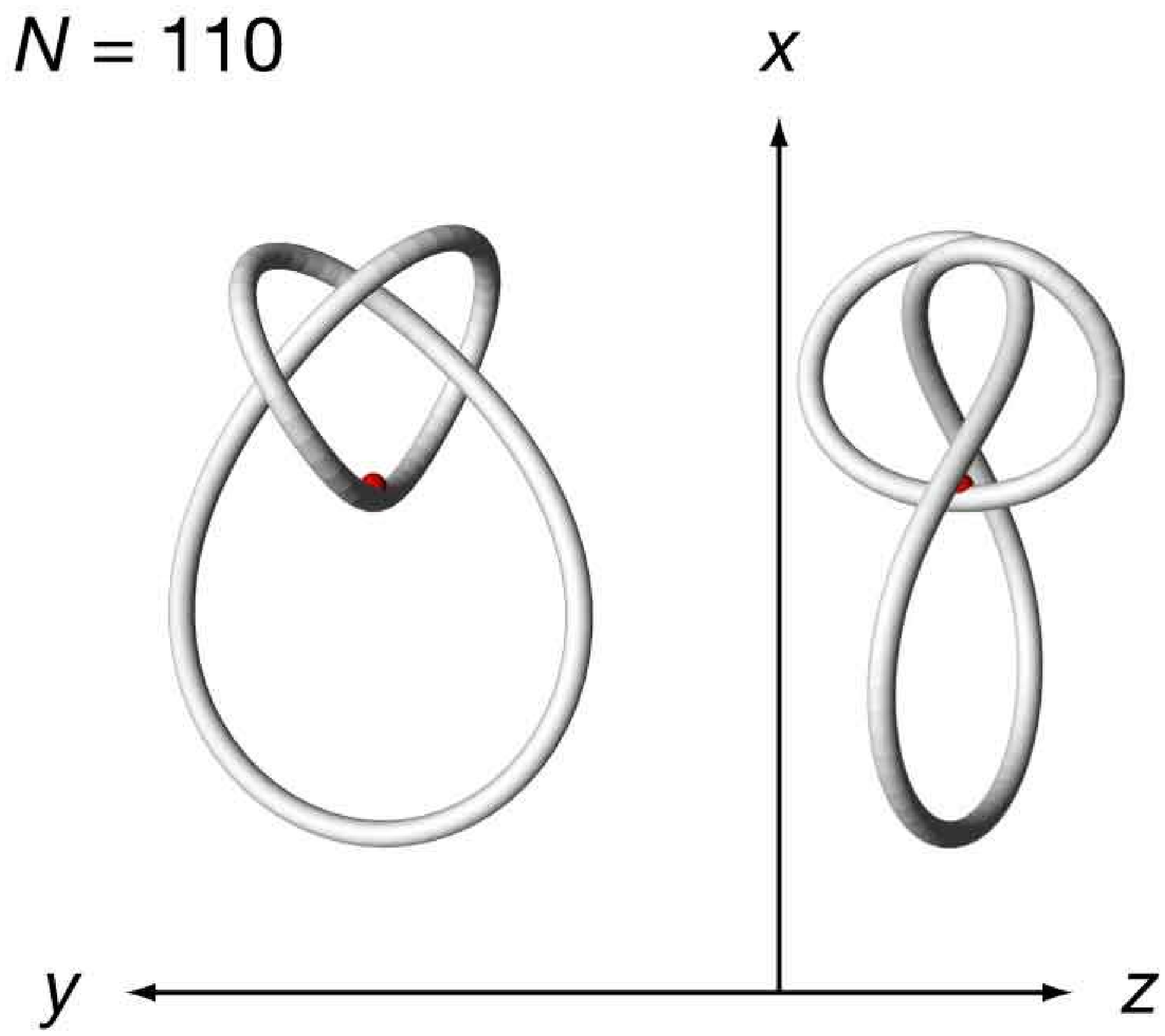}
	\includegraphics[height=5.5cm]{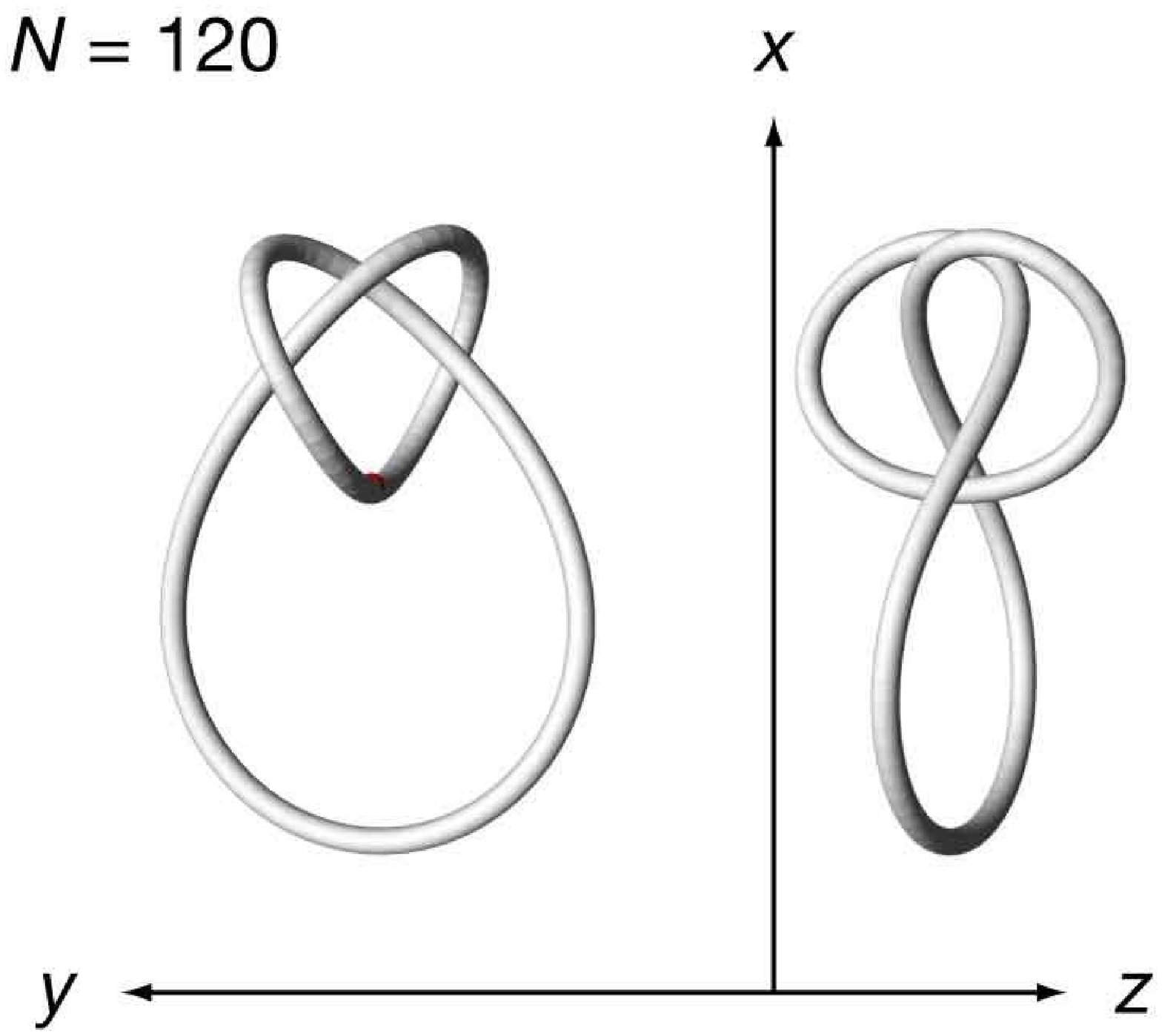}
	\\\medskip
	\includegraphics[height=5.5cm]{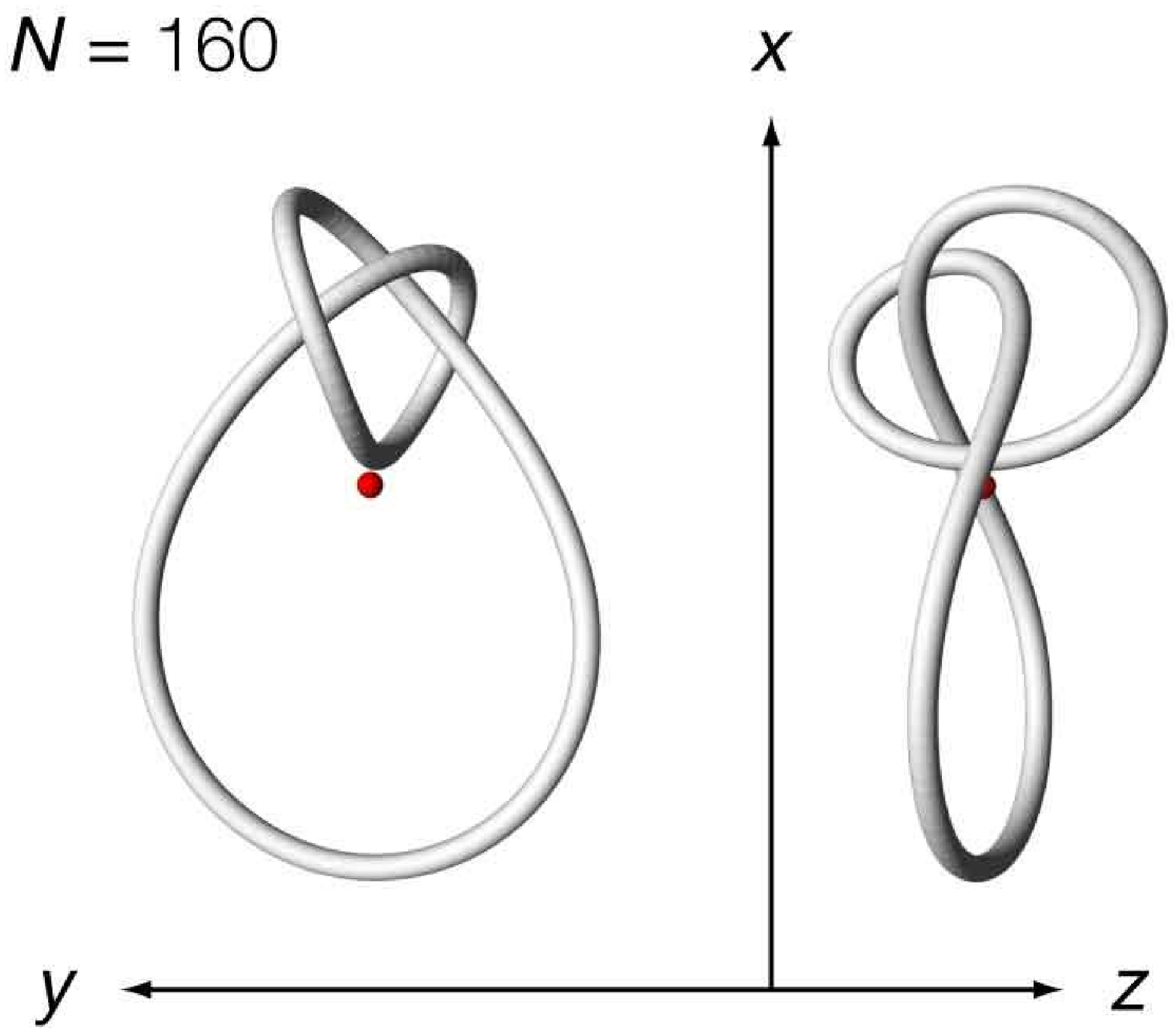}
	\includegraphics[height=5.5cm]{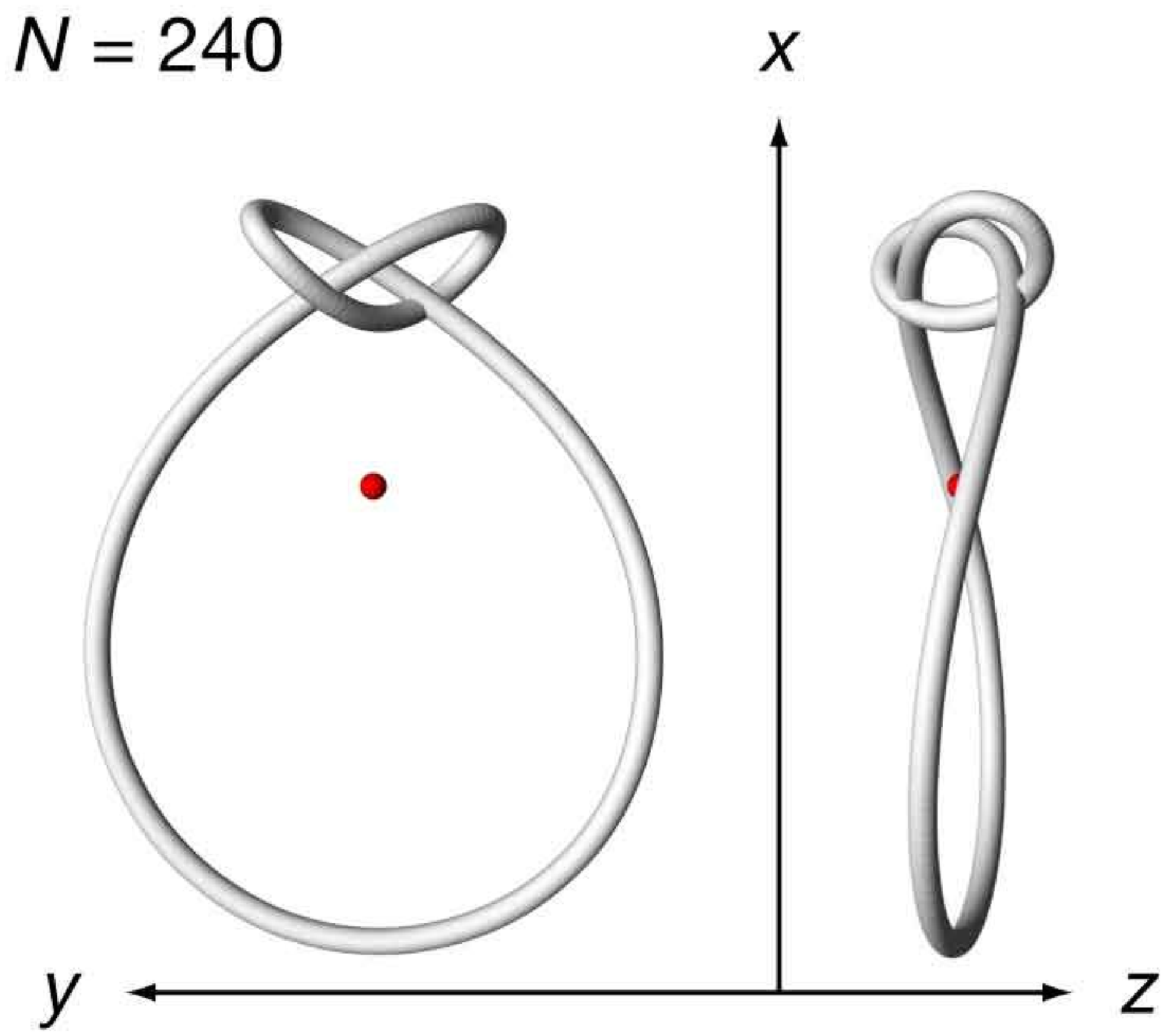}
	\caption{	
		The type-1 ASs
		of a single ring polymer with the trefoil knot
		for $N=30$, $40$, $60$, $80$, $110$, $120$, $160$ and $240$.
		The meaning of the cylinders and the spheres is
		the same as in Fig.\ \ref{fig:Average_Structure_Linear}.
		}
	\label{fig:Average_Structure_K=3_1 AS1}
\end{center}
\end{figure}

Figure \ref{fig:Average_Structure_K=3_1 AS1} shows
the type-1 ASs of single ring polymers
with the trefoil knot for
$N=30$, $40$, $60$, $80$, $110$, $120$, $160$ and $240$.
In the case of the trefoil knot,
the type-1 ASs do not have reflection symmetry.
This is because
a ring polymer with the trefoil knot has chirality
and
therefore
the mirror image of a sampled conformation $\mib{C}(m)$
is topologically different from $\mib{C}(m)$ and
does not appear in the ensemble of conformations
to which $\mib{C}(m)$ belongs.
Moreover,
the knot type of the type-1 AS is identical with
that of the original structure for each $N$.
It is clearly seen that
the knotted part is delocalized for small values of $N$
and is localized for large values of $N$.
The crossover segment number $N_{\rm x}^{{\rm av1}} \simeq 120$
from the delocalized state,
where the center of mass is inside of the knotted part,
to the localized state, 
where the center of mass is outside of the knotted part,
is consistent with the transition segment number $N_{\rm t}^{{\rm av0}}$ 
of the type-0 AS.
\section{Summary and Discussion}
\label{sec:Summary and Discussion}
In the present paper,
the two types of average structures are calculated
for a single linear polymer and single ring polymers
by Brownian dynamics simulations.
The average conformation vector, which specifies the average structure,
is self-consistently defined by eqs.\ (\ref{av0}) and (\ref{av1}):
It is the average of the sampled conformation vectors
each of which is rotated to minimize
its distance from the average conformation vector.
From a conformation vector $\mib{C}(m)$ of single homopolymers,
$K$ conformation vectors $\tilde{\mib{C}}(m,k)$ with $k=1, \ldots, K$
are generated
by changing the numbering of the segments,
where $K=2$ for linear homopolymers and $K=2N$
for ring homopolymers with $N$ segments.
In the calculation of
the type-0 average structure,
which has been used for heteropolymers,
the position of the $i$th segment of a sampled conformation
is fitted to  that of the $i$th segment of the average conformation.
Therefore, in the case of homopolymers,
the above-mentioned $K$ conformation vectors represent
different structures
and are used with the same statistical weight
in the calculation as shown in eq.\ (\ref{av0}).
Thus, the type-0 AS should be invariant
under the $K$ ways of changing the numbering of the segments.
For a single linear polymer,
the type-0 AS has a parabolic shape in the $x$-$y$ plane
with the C$_2$ symmetry about the $y$ axis,
where
the two symmetry operations of the group C$_2$
correspond to
the two ways of changing the numbering.
In the case of a single ring polymer of $N$ segments
with the trivial knot,
the type-0 AS is given by a regular polygon of $N$ sides
with the D$_N$ symmetry,
where the $2N$ symmetry operations of the group D$_N$
correspond to the $2N$ ways of changing the numbering.
\par
In the calculation of
the type-1 average structure,
which is proposed in the present paper,
differences in the numbering of the segments are neglected
and all the $K$ conformation vectors represent the same structure.
Therefore,
among the $K$ conformation vectors,
the only one conformation
$
\tilde{\mib{C}}(m,k_{\rm min}(m))
$,
which becomes
the closest to the average conformation vector
by the rotation
$
\mathcal{R}_{m,k_{\rm min}}(m)
$,
is used in the calculation of the type-1 AS
as shown in eq.\ (\ref{av1}).
Thus,
deviations from the symmetric form of the type-0 AS
in the sampled conformations
are conserved in the type-1 AS.
The type-1 AS of a single linear polymer
has a distorted parabolic form without the C$_2$ symmetry
in the $x$-$y$ plane.
For a ring polymer of $N$ segments with the trivial knot,
the type-1 AS forms a distorted polygon of $N$ sides
without the D$_N$ symmetry
in the $x$-$y$ plane.
It is shown that the mirror image of the average structure
calculated from conformations of a polymer
gives
the average structure
calculated from the mirror images of the conformations.
In the ensemble of conformations of
a linear polymer or a ring polymer with the trivial knot,
which has no chirality,
a conformation and its mirror image have the same statistical weight
and
the ensemble of the mirror images of the original conformations
is the same as the original ensemble.
Therefore
the average structure and its mirror image represent
the same structure.
Thus, the average structure
is given by a conformation in a plane or
a three-dimensional conformation with a plane of reflection symmetry.
All the average structures
of a single linear polymer and single ring polymers with the trivial
knot are planar and consistent with the above consideration.
\par
The type-0 AS of a single ring polymer of $N$ segments
with the trefoil knot
forms
a double loop on a regular polygon of $N/2$ sides for even $N \le 110$
and 
a single loop on
a regular polygon of $N$ sides for $N \ge 120$.
These structures are invariant
under
the symmetry operations corresponding to
the $2N$ ways of changing the numbering.
The transition from the double loop structure
to the single loop structure
occurs at
the transition segment number
$N_{\rm t}^{{\rm av0}}$ between $110$ and $120$.
This transition is considered to correspond to the
localization-delocalization transition of the knotted part.
Because a ring polymer with the trefoil knot is chiral,
its type-1 AS does not have reflection symmetry
and has the same knot type as the sampled conformations.
The knotted part of the type-1 AS
is expanded along the whole structure for small values of $N$ and 
localized to a part of the structure for large values of $N$.
The crossover from the delocalized state
to the localized state occurs around
the crossover segment number $N_{\rm x}^{{\rm av1}} \simeq 120$,
which is consistent with the value of $N_{\rm t}^{{\rm av0}}$.
The transition of the type-0 AS and 
the crossover of the type-1 AS
furnish strong evidence for the localization
of the knotted part predicted in the previous study,
where the transition segment number is between $120$ and $160$.%
\cite{Saka2008}
\par
It is demonstrated that
the analysis of the average structures is
useful for studying the localization of the knotted part
of single knotted ring polymers.
The present paper provides strong evidence for the knot localization
by directly observing the average structures in three dimensions.
In the case of single ring polymers with the trefoil knot, 
the transition of the type-0 AS
from the double loop structure to the single loop structure
is observed as the number of the segments $N$ is increased,
which is considered to correspond to
the localization-delocalization transition of the knotted part.
The trefoil knot is a $(3,2)$-torus knot.
Here, a $(p,q)$-torus knot is given by
a curve on the surface of a torus
which winds $p$ times around the center line of the torus
and revolves $q$ times along the center line of the torus.%
\cite{Lickorish}
In the case of the trefoil knot,
the curve revolves twice along the center line of the torus.
Therefore, it seems natural
that
the type-0 AS forms a double loop structure for small $N$
and that
the transition to a single loop structure occurs as $N$ is increased.
It is expected that the similar transitions occur
in the type-0 ASs of
single ring polymers with another torus knot,
such as $5_1$, $7_1$, or $9_1$.
It is interesting to study
how the type-0 ASs of single ring polymers with a nontorus knot
depends on $N$.
The type-1 ASs of single ring polymers with the trefoil knot
have the same knot type as the original conformations
and show the knot localization directly.
Single ring polymers with the trefoil knot have chirality.
Therefore, their type-1 ASs need not to have reflection symmetry.
Thus, single knotted ring polymers with chirality
have less constraints on the shapes of their type-1 ASs.
Because torus knots have chirality,
it is expected that
the type-1 ASs of
single ring polymers with another torus knot
show the knot localization in three dimensions.
It is interesting to study
what kind of structures are formed by
the type-1 ASs of single knotted ring polymers without chirality.
They must be
planar structures or three-dimensional structures with 
reflection symmetry.
The study of the average structures of single ring polymers
with the figure eight knot is very important,
because the figure eight knot
is the next simplest prime knot and
a nontorus knot without chirality.
The study in this direction is in progress.
\section*{Acknowledgments}
\label{sec:Acknowledgments}
\par
The authors are grateful to
Professor T.\ Deguchi, Dr.\ A.\ Mitsutake and Dr.\ K.\ Hagita
for their interest in the work and for useful discussions
and
Professor Y.\ Fujitani 
for providing computational resources.
This work was partially supported by the 21st Century COE Program;
Integrative Mathematical Sciences: Progress in Mathematics
Motivated by Social and Natural Sciences.

\end{document}